\documentclass[preprint,amsmath,amssymb,aps]{revtex4}
\usepackage{dcolumn}
\usepackage{bm}
\usepackage{graphicx}
\usepackage{subfigure}
\usepackage{color}
\usepackage[T1]{fontenc}
\usepackage[utf8]{inputenc}
\usepackage{lmodern}

\newcommand{\be}{\begin{eqnarray}}
\newcommand{\ee}{\end{eqnarray}}

\begin{document}
\title{Entanglement island and Page curve of Hawking radiation in rotating Kerr black holes}

\author{Liqiang Wang}
\thanks{Email: wangliqiangedu@outlook.com}

\author{Ran Li$^{1}$}
\thanks{Corresponding author, Email: liran@qfnu.edu.cn}

\affiliation{Department of Physics, Qufu Normal University, Qufu, Shandong 273165, China}

\numberwithin{equation}{section}

\begin{abstract}
   We initiate the study of the information paradox of rotating Kerr black holes by employing the recently proposed island rule. It is known that the scalar field theory near the Kerr black hole horizon can be reduced to the 2-dimensional effective theory. Working within the framework of the 2-dimensional effective theory and assuming the small angular momentum limit, we demonstrate that the entanglement entropy of Hawking radiation from the non-extremal Kerr black hole follows the Page curve and saturates the Bekenstein-Hawking entropy at late times. In addition, we also discuss the effect of the black hole rotation on the Page time and scrambling time. For the extreme Kerr black hole, the entanglement entropy at late times also approximates the Bekenstein-Hawking entropy of the extreme Kerr black hole. These results imply that entanglement islands can provide a semi-classical resolution of the information paradox for rotating Kerr black holes. 
\end{abstract}

\maketitle

\tableofcontents

\section{Introduction}\label{sect:intro}
\renewcommand{\theequation}{1.\arabic{equation}}
\setcounter{equation}{0}

According to Hawking's calculation \cite{hawking1975particle}, black holes can emit particles with a thermal spectrum of temperature proportional to the surface gravity of the event horizon—a process later known as Hawking radiation—causing black holes to evaporate and eventually vanish. This suggests that a black hole formed by the collapse of pure-state matter eventually evaporates into Hawking radiation in a maximally mixed state \cite{Wald:1975kc,Unruh:1976db}. The process of black hole evaporating violates the basic quantum principle of unitarity, which states that the final state resulting from the complete evaporation of a black hole that collapsed from zero-entropy pure-state matter must also be a pure state, rather than a maximally mixed state. Therefore, the information swallowed by the black hole cannot be recovered via Hawking radiation, giving rise to the well-known black hole information paradox \cite{hawking1976breakdown}. It is widely accepted that a deeper understanding of the black hole information paradox can provide meaningful insights into the theory of quantum gravity.

The black hole information paradox can also be explained from the perspective of von Neumann entropy or entanglement entropy. Hawking's calculation suggests that the entropy of Hawking radiation increases linearly during black hole evaporation \cite{hawking1976breakdown}. At late times, it will exceed the Bekenstein-Hawking entropy or the thermodynamic entropy of the black hole. If the black hole is initially in a pure state, the unitary evolution principle of quantum mechanics suggests that the entanglement entropy of the black hole should be equal to that of the radiation. If the entropy of radiation calculated by Hawking is treated as the entanglement entropy of radiation, the entanglement entropy of the black hole will exceed its thermodynamic entropy at late times \cite{Almheiri:2020cfm}. This contradicts the principle that the entanglement entropy of a system should be less than its thermodynamic entropy, based on the consideration of the number of degrees of freedom. If black hole evaporation is really an unitary process, the entanglement entropy of Hawking radiation would decrease at late times and eventually reach zero when the black hole is completely evaporated.

\begin{figure}
    \centering
    \includegraphics[width=8cm]{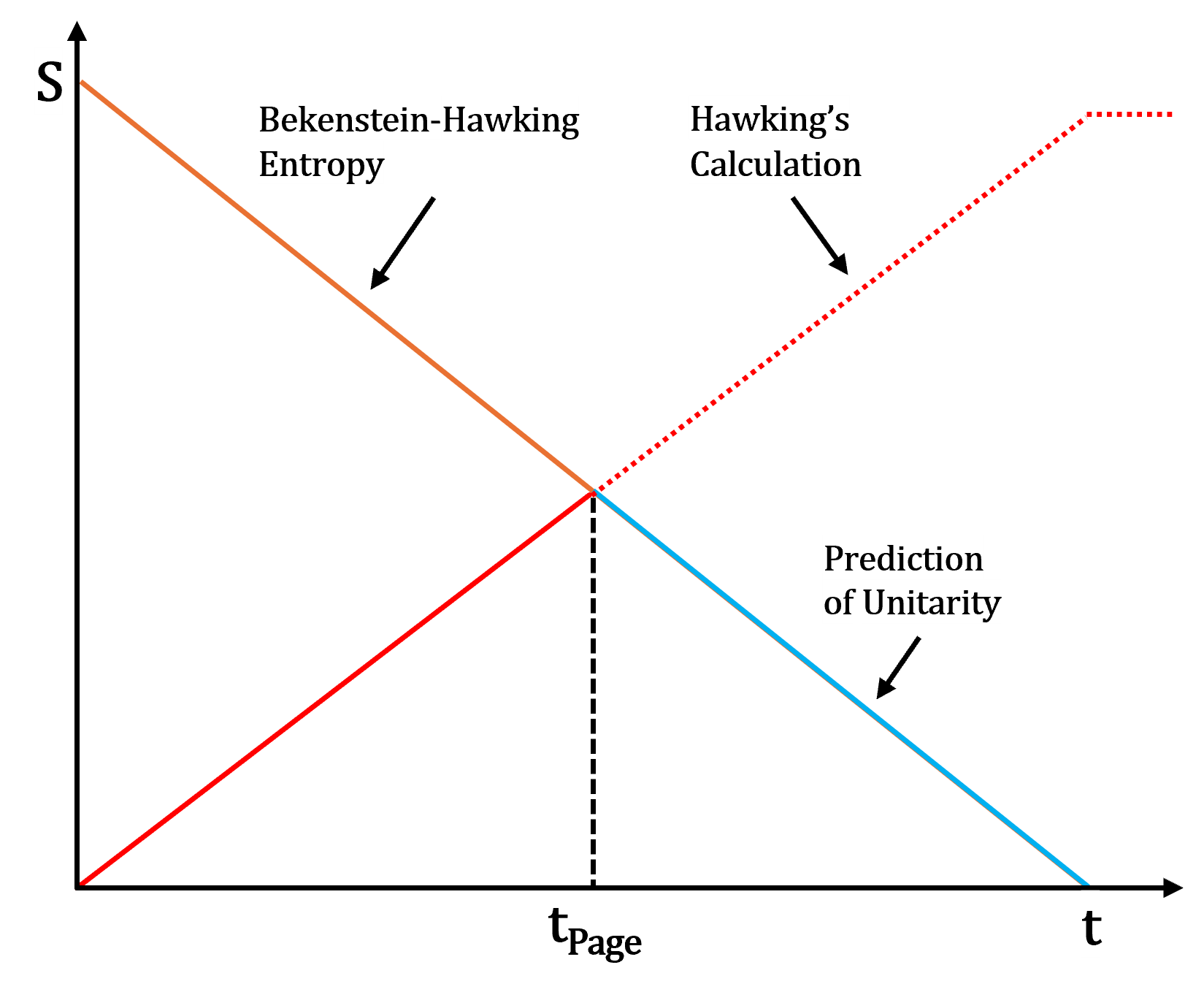}
    \caption{An illustration for the time evolution of the entanglement entropy of Hawking radiation. The red line and the red dashed line represent Hawking's results, the orange line represents the thermodynamic entropy of the black hole, and the red line and the blue line collectively constitute the Page curve.}
    \label{Page curve}
\end{figure}

Due to black hole evaporation is such a complex process, Page suggested that the entire system of the black hole and its radiation should be in a random pure state \cite{page1993information,page2013time}. It is also proved by Page \cite{Page:1993df} that the random pure state is almost maximally entangled as long as the number of degrees of freedom of one subsystem is much smaller than the number of degrees of freedom of the entire system. According to the Page's theorem, the entanglement entropy of radiation is roughly equal to its coarse-grained entropy at early times, and is approximately equal to the Bekenstein-Hawking entropy of the black hole at late times. As can be seen from Fig.(\ref{Page curve}), the entanglement entropy of the radiation increases from zero and reaches a maximum at the so-called Page time, and then decreases to zero. It is generally believed that being able to reproduce the Page curve in some particular theory means a resolution of the black hole information paradox \cite{harlow2016jerusalem}.

However, the assumption of unitary evolution of black hole evaporation was not justified until the discovery of AdS/CFT correspondence \cite{maldacena1999large}. The AdS/CFT correspondence, which states that the quantum gravity in AdS space is dual to a CFT on its boundary, provides the evidence of information conservation during black hole evaporation. This comes from the observation that the evaporation process of black hole in the AdS space can be described by a unitary process of the dual CFT on the boundary. Although the unitary evolution was criticized by the firewall paradox \cite{almheiri2013black}, it is generally believed that the black hole information paradox can be resolved in the framework of the AdS/CFT correspondence.

Recently, a breakthrough was made in calculating the entanglement entropy of Hawking radiation and reproducing the Page curve within the framework of the AdS/CFT correspondence \cite{Penington:2019npb,Almheiri:2019psf,Almheiri:2019hni,Penington:2019kki,Almheiri:2019qdq}. The advancement is mainly based on the pioneering work of Ryu-Takayanagi \cite{ryu2006holographic}, which relates the entanglement entropy of a boundary region to the area of a minimal surface in the bulk. When considering the quantum corrections, the RT proposal was generalized to the prescription of Quantum Extreme Surface (QES) \cite{hubeny2007covariant,lewkowycz2013generalized,barrella2013holographic,faulkner2013quantum,engelhardt2015quantum}. This provides a way to compute the entanglement entropy of the Hawking radiation, which is usually summarized as the island rule. The island rule specifies the entanglement entropy of Hawking radiation by the following formulas
\begin{align}
    S_{Rad}=\mathrm{min}\left\{ \mathrm{ext}\left[ S_{\text{gen}} \right] \right\}\;,
    \label{eq:general entropy}
\end{align}
\begin{align}
    S_{\text{gen}}=\frac{\mathrm{Area}\ \left[\partial I\right]}{4G_N}+S_{\mathrm{field}}\left(R\cup I\right)\
    \label{eq:island formula}\;,
\end{align}
where $G_N$ denotes the Newton constant, $R$ is the radiation
and $I$ is the entanglement island, $\partial I$ is the boundary of the island, and $S_{\mathrm{field}}$ denotes the entanglement entropy of the quantum fields (including gravitational field) in the radiation region $R$ and the island region $I$. The term $S_{\mathrm{gen}}$ in Eq.\eqref{eq:island formula} represents the generalized entropy. By extremizing the generalized entropy $S_{\mathrm{gen}}$, one can determine the location of the island. The minimum value is then chosen to be the entanglement entropy of the Hawking radiation if there are more than one extremum. The island formula can be derived from the Euclidean path integral on the replicated manifold \cite{Penington:2019kki,Almheiri:2019qdq}, even without the need for holography. It is shown that replica wormholes are the saddle points in the Euclidean path integral when the entanglement islands emerge. In addition, it is argued that the replica wormholes derivation of island formula is valid not only for the eternal black holes but also for the evaporating black holes \cite{Hartman:2020swn,Goto:2020wnk}. More recently, it is shown that the Ryu-Takayanagi conjecture in AdS/BCFT when the bulk is asymptotically AdS$_3$ can be derived from the replica path integral and the topological phase transition of the Ryu-Takayanagi surface corresponds to the formation of the replica wormhole on the Karch-Randall brane \cite{Geng:2024xpj}.

For the two dimensional evaporating black hole in JT gravity \cite{Almheiri:2019psf}, entanglement islands that emerge inside the horizon of the black hole after the Page time contribute nontrivialy to the entanglement entropy of the radiation, which allows the reproduction of the Page curve. A similar resolution of the information paradox would also be true for eternal black holes. It is shown that after the Page time, the entanglement entropy of the radiation from the eternal black hole is also modified by the emergence of an island. However, in this case, the island extends to the outer vicinity of the event horizon \cite{Almheiri:2019yqk}. As a result, the entanglement entropy of the radiation saturates the value of the Bekenstein-Hawking entropy of the black hole. Therefore, the island rule can also provide a resolution of information paradox for the eternal black holes.

The proposal of island rule was initially inspired by studies of two-dimensional black holes in JT gravity. It has since been applied to investigate a wide class of black hole spacetimes across different dimensions, various asymptotic behaviors, and higher derivative gravity theories \cite{Chen:2019uhq,Almheiri:2019psy,Gautason:2020tmk,Anegawa:2020ezn,Hashimoto:2020cas,Hollowood:2020cou,Krishnan:2020oun,Alishahiha:2020qza,Chen:2020uac,Geng:2020qvw,Li:2020ceg,Chandrasekaran:2020qtn,Bak:2020enw,Krishnan:2020fer,Chen:2020jvn,Hartman:2020khs,Balasubramanian:2020xqf,Balasubramanian:2020coy,Sybesma:2020fxg,Chen:2020hmv,Ling:2020laa,Hernandez:2020nem,Matsuo:2020ypv,Karananas:2020fwx,Wang:2021woy,Geng:2021wcq,Fallows:2021sge,Bhattacharya:2021jrn,Kim:2021gzd,Wang:2021mqq,Geng:2021iyq,Uhlemann:2021nhu,Li:2021lfo,Chu:2021gdb,Lu:2021gmv,Yu:2021cgi,Ahn:2021chg,Aguilar-Gutierrez:2021bns,Kames-King:2021etp,Cao:2021ujs,Saha:2021ohr,Azarnia:2021uch,Okuyama:2021bqg,Omidi:2021opl,Yu:2021rfg,Gan:2022jay,Seo:2022ezk,Azarnia:2022kmp,Tian:2022pso,Afrasiar:2022ebi,Anand:2022mla,Djordjevic:2022qdk,Goswami:2022ylc,Yu:2022xlh,Hu:2022zgy,Lu:2022tmt,BenDayan:2022nmb,Baek:2022ozg,Guo:2023gfa,Parvizi:2023foz,Wu:2023uyb,Li:2023fly,RoyChowdhury:2023eol,Wang:2023eyb,Tong:2023nvi,Yu:2023whl,Chou:2023adi,Chang:2023gkt,Matsuo:2023cmb,Anand:2023ozw,Li:2023zgy,Xu:2023fad,Ageev:2023hxe,Bousso:2023kdj,Jeong:2023lkc,Yadav:2022fmo,Yadav:2022jib,Yadav:2023sdg,Yadav:2022mnv,RoyChowdhury:2022awr,Bhattacharya:2021dnd,Bhattacharya:2021nqj}. The island rule has been shown to provide the resolution of the information paradox for these black holes by reproducing the Page curves. However, the proposal of the island rule does not mean a complete solution to the information paradox, as many aspects of its exact physical meaning still need clarification. In addition, among the works mentioned above, the rotating black holes are rarely considered due to the complexity of their spacetime geometry. Rotating black hole, also known as Kerr black hole, is a type of black hole that possesses mass and angular momentum. It is characterized by a ring-shaped singularity and an ergoregion outside of the horizon. Kerr black holes are probably the commonest in the realistic four-dimensional universe. Therefore, it is important to address the information paradox in this type of rotating black hole.

In four or high-dimensional spacetime, it is hard to calculate the entanglement entropy of the conformal field. Usually, the two-dimensional s-wave approximation \cite{Hashimoto:2020cas} is employed to calculate the entanglement entropy of the matter fields. On the other side, it is well known that the scalar field theory near the Kerr black hole horizon can be reduced to the 2-dimensional effective theory. We reduce the scalar field theory in 4-dimensional Kerr spacetime to a 2-dimensional field theory following the method in \cite{murata2006hawking}. In the present work, we will work with the 2-dimensional effective theory of the four dimensional Kerr black hole. Due to the nonlinearity of Einstein's equations in higher-dimensional spacetimes, we will ignore the backreaction of the matter field on the spacetime to simplify this analysis.

Note that the Kerr black hole has a significant effect caused by rotation, superradiance \cite{1971JETPL..14..180Z,1974ApJ...193..443T}, which may result in the instability of the spacetime \cite{Press:1972zz}. Both Hawking evaporating and superradiance process contribute the radiation outside of the Kerr black hole. It is a rather complicated case when taking the both effects into account. We will consider the case that Hawking radiation dominates over the superradiance by assuming the mass of the Kerr black hole is much larger than its angular momentum.

By assuming the small angular momentum limit, we demonstrate that the entanglement entropy of Hawking radiation from the non-extremal Kerr black hole saturates the coarse-grained entropy of the black hole and follows the Page curve. In addition, we discuss the effect of the black hole rotation on the Page time and scrambling time. For the extreme Kerr black hole, the entanglement entropy at late times also approximates the coarse-grained entropy of the extreme Kerr black hole. These results imply that entanglement islands can provide a semi-classical resolution of the information paradox for the rotating Kerr black holes.

This paper is arranged as follows. In Section \ref{sect:Reduction}, we briefly review the basic facts about the Kerr black hole and its reduced 2-dimensional effective theory. In Section \ref{sect:non-extremal_Kerr}, we calculate the entanglement entropy for Hawking radiation in a non-extremal eternal Kerr black hole and reproduce the Page curve. Additionally, we discuss the Page time and scrambling time for non-extremal Kerr black holes. In Section \ref{sec:extremal_BHs}, we consider the extremal Kerr black hole case and show that the entanglement entropy of Hawking radiation in the late stage approaches the Bekenstein–Hawking entropy. Section \ref{sec:con_disc} provides discussion and conclusions.

\section{Kerr black hole and its 2-dimensional reduction}\label{sect:Reduction}
\renewcommand{\theequation}{2.\arabic{equation}}
\setcounter{equation}{0}
\subsection{Kerr black holes}

The metric of the Kerr black holes in Boyer-Lindquist coordinates is given by
\begin{align}\label{Kerr_metric}
    \mathrm{d}s^2 = & -\frac{\Delta - a^2 \sin^2{\theta}}{\Sigma} \mathrm{d}t^2 - \frac{2a\sin^2{\theta}(r^2 + a^2 - \Delta)}{\Sigma} \mathrm{d}t \mathrm{d}\phi \notag           \\
                    & +\frac{(r^2 + a^2)^2 - \Delta a^2 \sin^2{\theta}}{\Sigma} \sin^2{\theta} \mathrm{d}\phi^2 + \frac{\Sigma}{\Delta} \mathrm{d}r^2 + \Sigma \mathrm{d}\theta^2\;,
\end{align}
where
\begin{align}
    \Sigma = r^2 + a^2 \cos^2{\theta}, \quad \Delta = r^2 - 2Mr + a^2 = (r - r_+)(r - r_-).
\end{align}
Here $M$ is the mass of the Kerr black hole and $a$ is its rotation parameter. When $M>a$, there are two solutions $r_+$ and $r_-$ to the equation $\Delta=0$. The metric \eqref{Kerr_metric} describes the non-extremal Kerr black holes and $r_+$ and $r_-$ represent the outer and inner horizons, respectively. When $M=a$, the inner and outer horizons coincide and the metric \eqref{Kerr_metric} describes the extremal Kerr black holes.

The surface gravity $\kappa$ at the outer horizon can be written as
\begin{align}
    \kappa =\sqrt{-\frac{1}{2}\left( \nabla ^a\xi _b \right) \left( \nabla _a\xi ^b \right)}=\frac{r_+-r_-}{2\left( {r_+}^2+a^2 \right)}\;,
\end{align}
where $\nabla_a$ is the covariant derivative operator, and $\xi^a$ is a non-null Killing vector at the event horizon. The Hawking temperature is given by 
\begin{eqnarray}
    T_H=\frac{\kappa}{2\pi}=\frac{r_+-r_-}{4\pi\left( {r_+}^2+a^2 \right)}\;.
\end{eqnarray}
The Bekenstein-Hawking entropy is given by 
\begin{eqnarray}
    S_{BH}=\frac{\pi\left(r_+^2+a^2\right)}{G_N}\;.
\end{eqnarray}
The angular velocity of the event horizon is given by 
\begin{eqnarray}
    \Omega_H=\frac{a}{r_+^2+a^2}\;.
\end{eqnarray}

In particular, for extreme black holes with the equal inner and outer horizons, the surface gravity $\kappa=0$, which results in the vanishing Hawking temperature.

\subsection{Dimensional reduction of Kerr metric}

In this subsection, we will consider the action of a scalar field in the background of four dimensional Kerr black hole and follow the line of \cite{murata2006hawking} to show how to obtain the two dimensional effective theory. The procedure also allows us get the reduced two-dimensional metric.

The action for the scalar field $\varphi$ in 4-dimensional Kerr spacetime is given as
\begin{align}
    S\left[ \varphi \right] =\frac{1}{2}\int_V{\mathrm{d}^4x\sqrt{-g}\varphi \nabla ^2\varphi}\;,
\end{align}
where $g$ is the determinant of the 4-dimensional Kerr metric. In terms of the metric \eqref{Kerr_metric}, by taking the limit $r\rightarrow r_+$, the action can be approximated as 
\begin{align}
    S\left[ \varphi \right] =\int_V{\mathrm{d}^4x\varphi \sin \theta \left[ -\frac{\left( r_++a^2 \right) ^2}{\Delta}{\partial _t}^2-\frac{2a\left( {r_+}^2+a^2 \right)}{\Delta}\partial _t\partial _{\phi}+\partial _r\Delta \partial _r-\frac{a^2}{\Delta}{\partial _{\phi}}^2 \right] \varphi}
\end{align}
To transform the Boyer-Lindquist coordinates of the Kerr black hole to the locally nonrotating coordinate system, we introduce the following coordinate transformation
\begin{align}
    \begin{cases}
        \psi =  \phi - \Omega_H t, \\
        \xi = t.
    \end{cases}
\end{align}
By substituting the coordinate transformation into the action of the scalar field, one can get
\begin{align}
    S\left[ \varphi \right] =\frac{a}{2\Omega _H}\int_V{\mathrm{d}^4x\varphi \sin \theta \left[ -\frac{1}{f\left( r \right)}{\partial _{\xi}}^2+\partial _rf\left( r \right) \partial _r \right] \varphi}\;,
\end{align}
where
\begin{align}
    f(r)=\frac{\Omega_H \Delta}{a}\;.
\end{align}
It is obvious that the angular derivative terms in the action disappears completely. Then, by expanding the scalar field $\varphi$ in terms of spherical harmonics as $\varphi(x)=\sum_{l,m} \varphi_{lm}(\xi,r) Y_{lm}(\theta,\psi) $, we can get the effective two dimensional action as 
\begin{align}
    S\left[ \varphi \right] =\frac{a}{2\Omega _H}\sum_{l,m}{\int_V{\mathrm{d}\xi \mathrm{d}{r\varphi _l}_m\left\{ \partial _{\xi}\left[ -\frac{1}{f\left( r \right)}\partial _{\xi}{\varphi _l}_m \right] +\partial _r\left[ f\left( r \right) \partial _r{\varphi _l}_m \right] \right\}}}\;.
\end{align}
The effective 2-dimensional metric can be obtained from the above action as:
\begin{align}
    \mathrm{d}s^2=-f\left( r \right) \mathrm{d}\xi ^2+\frac{1}{f\left( r \right)}\mathrm{d}r^2
    \label{eq:effective 2-dimensional metric}\;.
\end{align}

Thus, we have shown that the four-dimensional scalar field can be reduced to a two-dimensional one by taking the near-horizon limit. Close to the outer horizon $r_+$, the non-extremal Kerr spacetime can be approximated by Rindler space. For extremal Kerr spacetime, where $r_+ = r_-$, the near-horizon geometry can be approximated by $\text{AdS}_2$ geometry \cite{Bardeen:1999px}.

In the following chapters, we will work with the two dimensional effective geometry to study the information paradox of the non-extremal and the extremal Kerr black holes. We will calculate the entanglement entropy of Hawking radiation in a 4-dimensional Kerr spacetime by utilizing an effective 2-dimensional spacetime framework. For non-extremal Kerr black holes, both Hawking radiation and superradiation processes contribute to the emitted radiation. To ensure that Hawking radiation dominates over superradiation, the black hole's mass must be substantially greater than its rotational energy. Therefore, we will work in the small angular momentum limit, i.e. $a\ll M$.

Furthermore, we assume the radiation is described by the conformal field with the central charge $c$. To neglect backreaction effects, we assume the black hole is macroscopic, with the central charge $c$ satisfying the condition $1\ll c \ll M$. The radiation dynamics can be semi-classically described as a 2-dimensional conformal field theory (CFT). By disregarding the grey-body factor, we can approximate the entanglement entropy in the 4-dimensional Kerr spacetime using the analysis of the 2-dimensional CFT.

In principle, any number and any size of islands can exist. However, when there is more than one island, the situation becomes more complex. For simplicity, we restrict our discussion to cases with either no island or only one island. We will confirm that this restriction is sufficient to solve the information paradox and to provide a reasonable Page curve.

\subsection{Tortoise coordinates of non-extremal case}

According to the reduced 2-dimensional metric given in Eq.\eqref{eq:effective 2-dimensional metric}, the corresponding tortoise coordinate $r^\ast\left(r\right)$ is given by:
\begin{align}
    r^*\left( r \right) =\int{f^{-1}\left( r \right) \mathrm{d}r}=\int{\frac{{r_+}^2+a^2}{\left( r-r_+ \right) \left( r-r_- \right)}}\mathrm{d}r=\frac{{r_+}^2+a^2}{r_+-r_-}\ln \left| \frac{r-r_+}{r-r_-} \right|+C\;,
\end{align}
where $C$ is an integration constant. For simplicity, we can set $C=0$. We then perform the following coordinate transformations: 
\begin{eqnarray}
    u=\xi-r^\ast\;,\;\;\;v=\xi+r^\ast\;.
\end{eqnarray}
The metric in the $u$ and $v$ coordinates is given by
\begin{align}
    \mathrm{d}s^2=-f\left( r \right) \mathrm{d}u\mathrm{d}v=-\frac{\left( r-r_+ \right) \left( r-r_- \right)}{{r_+}^2+a^2}\mathrm{d}u\mathrm{d}v
\end{align}

\begin{figure}
    \centering
    \includegraphics[width=8cm]{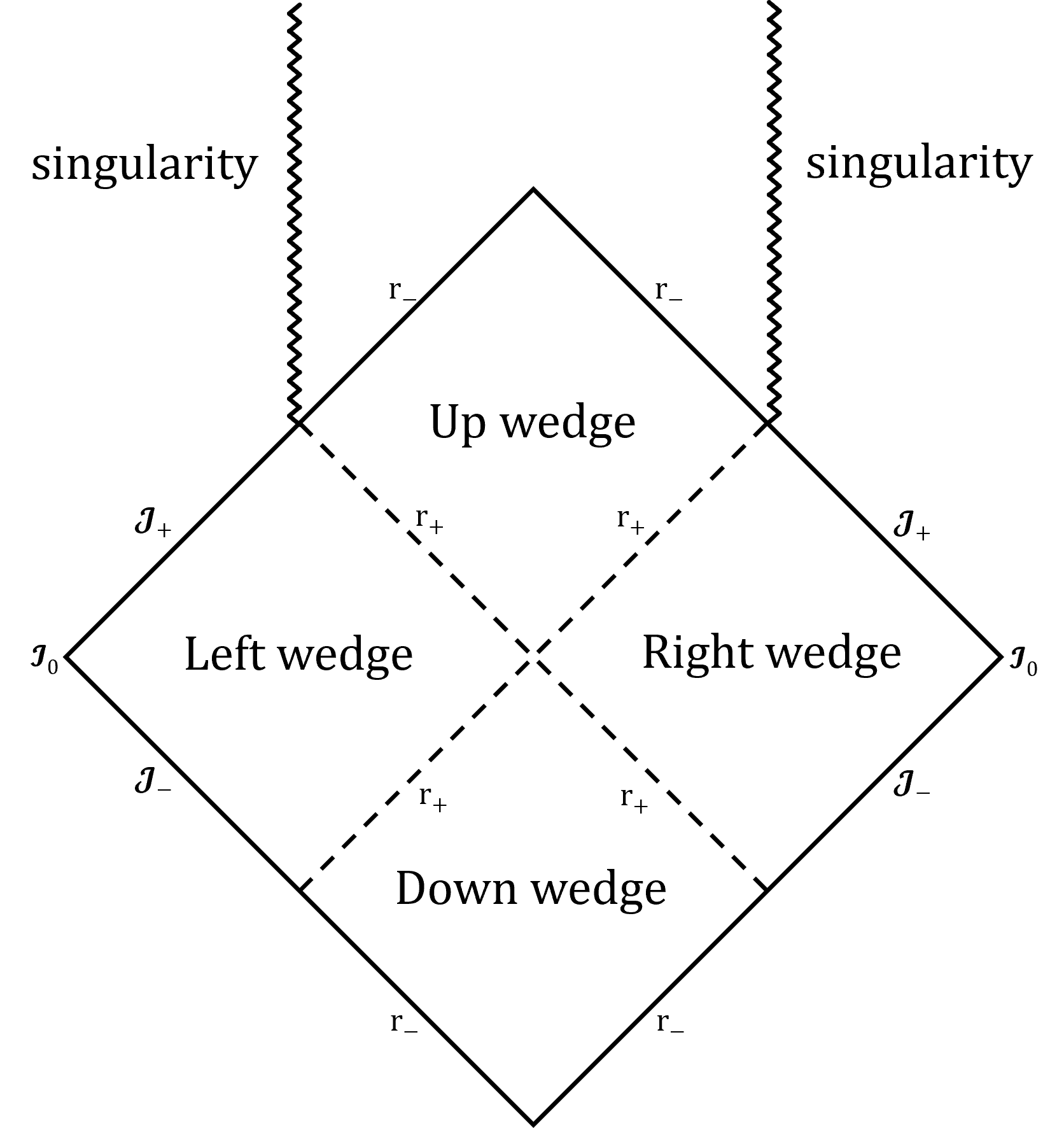}
    \caption{Penrose diagram of the non-extremal Kerr black hole. Here, $r_-$ represents the inner horizon of the black hole, $r_+$ represents the outer horizon, $\mathcal{J}_-$ is past null infinity, $\mathcal{J}_+$ is future null infinity, and $\mathcal{I}_0$ is spacelike infinity.}
    \label{The penrose diagram of nonextreme Kerr black hole}
\end{figure}

In order to eliminate coordinate singularities, the Kruskal coordinates is defined as
\begin{align}
    \text{Up wedge:} \quad V    & = \frac{e^{\kappa v}}{\kappa}, \quad U = \frac{e^{-\kappa u}}{\kappa} \label{eq:up_wedge of nonextreme case}     \\
    \text{Down wedge:} \quad V  & = -\frac{e^{\kappa v}}{\kappa}, \quad U = -\frac{e^{-\kappa u}}{\kappa} \label{eq:down_wedge of nonextreme case} \\
    \text{Right wedge:} \quad V & = \frac{e^{\kappa v}}{\kappa}, \quad U = -\frac{e^{-\kappa u}}{\kappa} \label{eq:right_wedge of nonextreme case} \\
    \text{Left wedge:} \quad V  & = -\frac{e^{\kappa v}}{\kappa}, \quad U = \frac{e^{-\kappa u}}{\kappa} \label{eq:left_wedge of nonextreme case}
\end{align}
Note that the spacetime has been maximally extended. The corresponding Penrose diagram for the Kerr black hole in Kruskal coordinates is shown in Figure \ref{The penrose diagram of nonextreme Kerr black hole}. The metric of the two-dimensional effective spacetime in $U$ and $V$ coordinates is given by
\begin{align}
    \mathrm{d}s^2=-\Omega ^2\mathrm{d}U\mathrm{d}V\;,
    \label{eq:metric in Kruskal coordinate}
\end{align}
where the conformal factor $\Omega$ is given by:
\begin{align}
    \Omega =\sqrt{f\left( r \right)}\exp \left( -\kappa r^* \right)\;.
\end{align}
This procedure shows that the spacetime is conformally flat, which means that in the Kruskal coordinates, the geodesic distance $d\left(x_1, x_2\right)$ between two spacetime points $x_1$ and $x_2$ is given by
\begin{align}
    d\left( x_1,x_2 \right) =\Omega \left( x_1 \right) \Omega \left( x_2 \right) \left[ U\left( x_2 \right) -U\left( x_1 \right) \right] \left[ V\left( x_1 \right) -V\left( x_2 \right) \right]\;.
    \label{eq:the geodesic distance between two points}
\end{align}

\subsection{Tortoise coordinates for the extremal case}

For extremal black holes, where $r_+=r_-=M=a$, we use $r_h$ to denote the horizon radius. According to the reduced 2-dimensional metric given in Eq.\eqref{eq:effective 2-dimensional metric}, the tortoise coordinate $r^\ast\left(r\right)$ is defined as
\begin{align}
    r^*\left( r \right) =-\left| \int{f^{-1}\left( r \right) \mathrm{d}r} \right|=-\left| \int{\frac{{r_h}^2+a^2}{\left( r-r_h \right) ^2}\mathrm{d}r} \right|=-\frac{2{r_h}^2}{\left| r-r_h \right|}+C\;.
\end{align}
Set $C=0$ for simplicity. In the extreme case, the reduced two-dimensional metric in $u$ and $v$ coordinates is given by
\begin{align}
    \mathrm{d}s^2=-f\left( r \right) \mathrm{d}u\mathrm{d}v=-\frac{\left( r-r_h \right) ^2}{2{r_h}^2}\mathrm{d}u\mathrm{d}v\;.
\end{align}

\begin{figure}
    \centering
    \includegraphics[width=8cm]{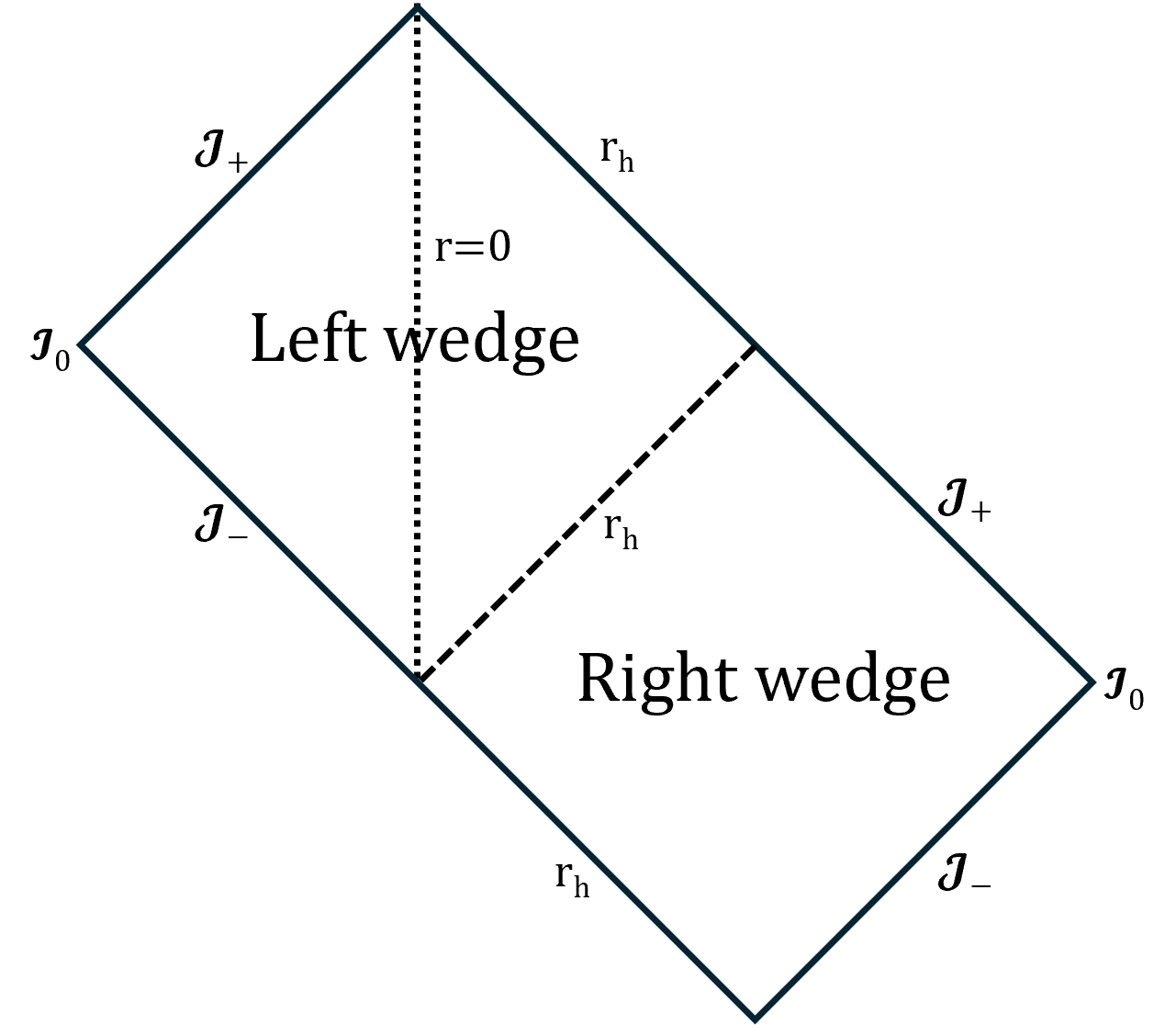}\\
    \caption{A Penrose diagram of the region of spacetime covered by Kruskal coordinates of an extremal Kerr black hole. Here, $r_h$ represents the horizon of the black hole, $\mathcal{J}_-$ is past null infinity, $\mathcal{J}_+$ is future null infinity, and $\mathcal{I}_0$ is spacelike infinity.}
    \label{The penrose diagram of extreme Kerr black hole}
\end{figure}

The Kruskal coordinates is defined as follows:
\begin{align}
    \text{Left wedge:} \quad  & V = -\frac{e^{\kappa v}}{\kappa}, \quad U = \frac{e^{-\kappa u}}{\kappa} \label{eq:left_wedge of extreme case}  \\
    \text{Right wedge:} \quad & V = \frac{e^{\kappa v}}{\kappa}, \quad U = -\frac{e^{-\kappa u}}{\kappa} \label{eq:right_wedge of extreme case}
\end{align}
where the parameter $\kappa$ is given by $\kappa = 1/r_h$. Note that $\kappa$ is not the surface gravity of the extremal black hole. The Penrose diagram for the extremal Kerr black hole is shown in Figure \ref{The penrose diagram of extreme Kerr black hole}. The metric for the extremal Kerr black hole in the Kruskal coordinates has the same form of the non-extremal Kerr black hole as given in Eq.\eqref{eq:metric in Kruskal coordinate}. The extremal black hole spacetime is also conformally flat, similar to the non-extremal Kerr black hole. The geodesic distance can also be calculated from Eq.\eqref{eq:the geodesic distance between two points}.

\section{Page curve for non-extremal Kerr black hole}
\label{sect:non-extremal_Kerr}

\renewcommand{\theequation}{3.\arabic{equation}}
\setcounter{equation}{0}

\subsection{Entanglement Entropy without island}
\label{sect:without Island}

\begin{figure}
    \centering
    \includegraphics[width=8cm]{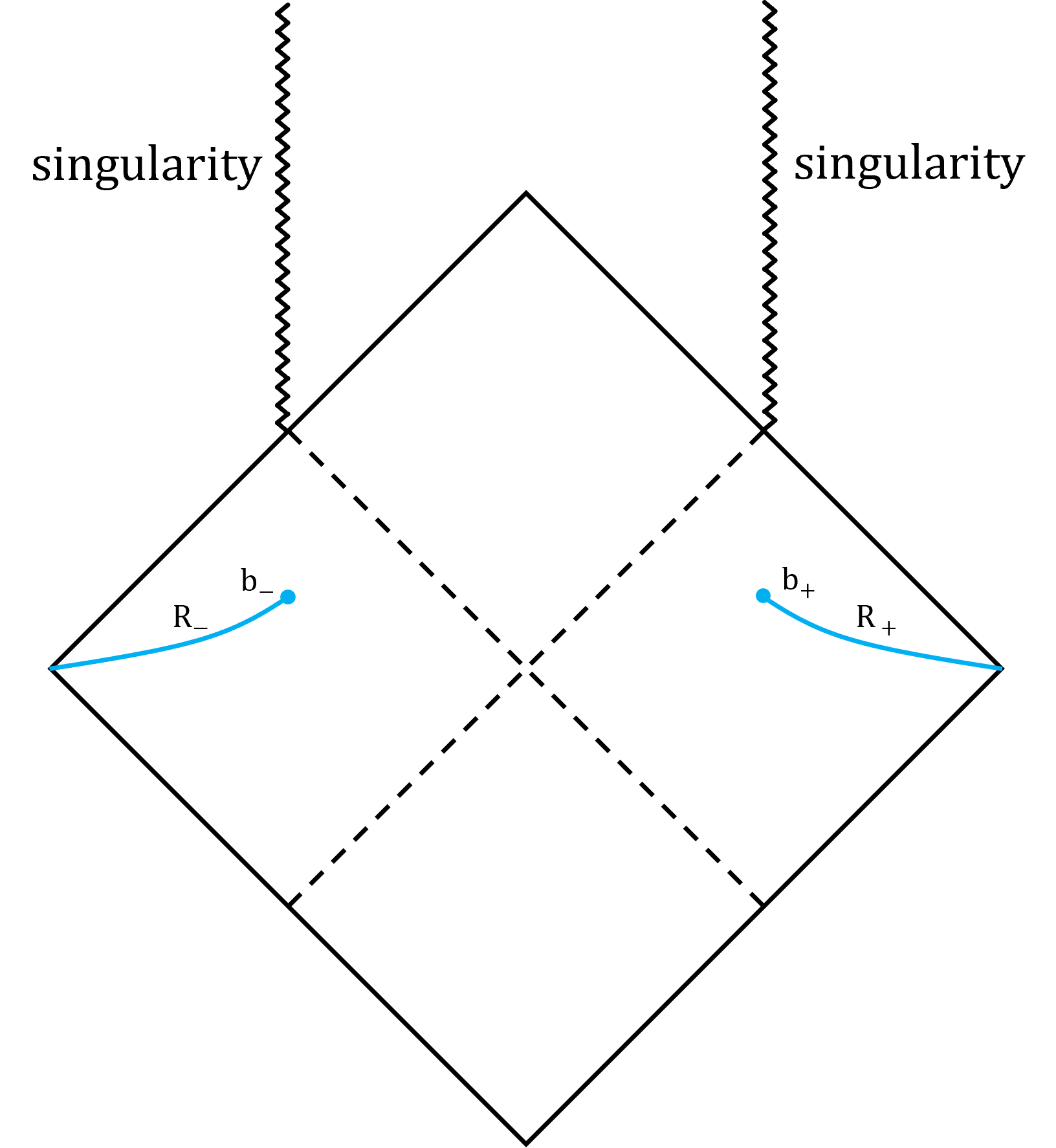}\\
    \caption{Penrose diagram of an eternal non-extremal Kerr black hole without island. The Hawking radiation is assumed to be located in the regions $R_\pm$, with boundaries $b_\pm$.}
    \label{Penrose diagram without island}
\end{figure}

Without considering the contributions of the island, we will compute the entanglement entropy of the Hawking radiation from a non-extremal Kerr black hole in the late stage of evaporation. The Hawking radiation is collected in the regions $R_\pm$, as shown in Figure \ref{Penrose diagram without island}. We assume this region is far from the horizon of the black hole, and the dynamics of the Hawking radiation can be approximated by a conformal field theory in flat spacetime. Without island, as shown in Figure \ref{Penrose diagram without island}, the end-points of the entanglement wedge of the radiation are the boundaries $b_+$ and $b_-$ of the right region $R_+$ and the left region $R_-$, respectively.

Assuming there is no island forming in the late stage, we should calculate the entanglement entropy of the CFT for the region outside $\left[b_-, b_+\right]$. If the whole system is in a pure state at $t=0$, then the entropy of the region $\left[b_-, b_+\right]$ is equal to that of the outside region. Therefore, the entanglement entropy of the radiation region is given by:
\begin{align}
    S_R=\frac{c}{3}\ln d\left( b_+,b_- \right)\;,
    \label{eq:von Neumann entropy of the radiation region 1}
\end{align}
where $d(b_+, b_-)$ is the geodesic distance between the boundary points $b_+$ and $b_-$. For $b_+$ we have $\left( \xi, r \right) = \left( t_b, r_b \right)$, and for $b_-$ we have $\left( \xi, r \right) = \left( -t_b + i\beta /2, r_b \right)$ with $\beta=\frac{2\pi}{\kappa}$ being the inverse Hawking temperature.

From equations (\ref{eq:metric in Kruskal coordinate}), (\ref{eq:the geodesic distance between two points}), and (\ref{eq:von Neumann entropy of the radiation region 1}), the entanglement entropy of the Hawking radiation in Kerr spacetime takes the form of:
\begin{align}
    S_R=\frac{c}{6}\ln \left\{ \Omega \left( b_+ \right) \Omega \left( b_- \right) \left[ U\left( b_- \right) -U\left( b_+ \right) \right] \left[ V\left( b_+ \right) -V\left( b_- \right) \right] \right\}\;.
\end{align}
When we substitute the coordinates of $b_+$ and $b_-$, the entanglement entropy of the Hawking radiation can be expressed as:
\begin{align}
    S=\frac{c}{3}\ln \left[ \frac{2\sqrt{f\left( r_b \right)}}{\kappa}\cosh \left( \kappa t_b \right) \right]\;.
    \label{eq:von Neumann entropy of the Hawking radiation 2}
\end{align}
Without island, Eq.(\ref{eq:von Neumann entropy of the Hawking radiation 2}) shows the entanglement entropy of the Hawking radiation as a function of time. At the late stage of evaporation, the entropy can be approximated as:
\begin{align}
    S^{\left( late \right)}\approx \frac{c}{3}\ln \left[ 4\sqrt{\frac{\left( r_b-r_+ \right) \left( r_b-r_- \right) \left( {r_+}^2+a^2 \right)}{\left( r_+-r_- \right) ^2}} \right] +\frac{c}{3}\kappa t_b\approx \frac{c}{3}\kappa t_b\;.
\end{align}

Consequently, without the island, the information that falls into the black hole cannot escape, and the entanglement entropy grows linearly with time. The entanglement entropy of the Hawking radiation will increase linearly towards infinity and eventually exceed the Bekenstein-Hawking entropy of the Kerr black hole. Assuming the eternal Kerr black hole is sustained by feeding it pure state matter, the total von Neumann entropy of the black hole will not change. For an eternal Kerr black hole, the maximum entanglement entropy of the Hawking radiation will be twice the Bekenstein-Hawking entropy of a one-sided Kerr black hole, taking into account contributions from both the left and right wedges in the conformal diagram. Therefore, no-island computation leads to the information paradox for the eternal Kerr black holes. The next section will show how the inclusion of the island can resolve the black hole information problem and yield the expected Page curve for an eternal non-extremal Kerr black hole.

\subsection{Entanglement Entropy with Island}\label{sect:with Island}

\begin{figure}
    \centering
    \includegraphics[width=8cm]{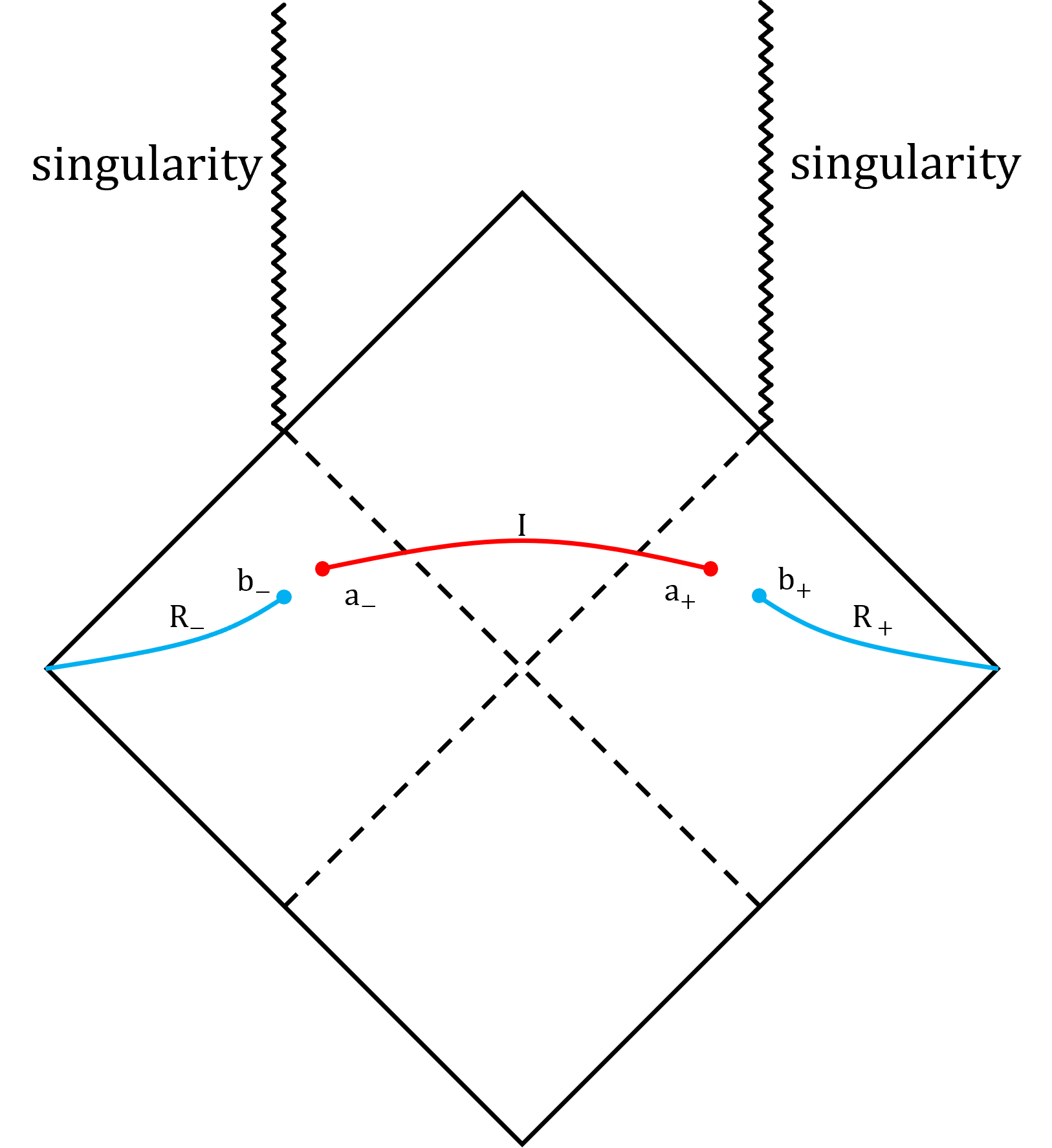}\\
    \caption{Penrose diagram of an eternal non-extremal Kerr black hole with the island. The Hawking radiation is assumed to be located in the regions $R_\pm$, with boundaries $b_\pm$. The island region, denoted as $I$, has boundaries $a_\pm$.}
    \label{Penrose diagram with island}
\end{figure}

Now, we calculate the entanglement entropy of the radiation by consider the contribution from the island. As shown in Figure \ref{Penrose diagram with island}, an island region $I$ with boundaries $a_+$ and $a_-$ is introduced in the Penrose diagram. We will follow the island rule, which is outlined in Eq.(\ref{eq:general entropy}) and Eq.(\ref{eq:island formula}), to study the information paradox of Kerr black hole.

The von Neumann entropy of the matter fields in the region $R \cup I$ is given by:
\begin{align}
    S_{field}\left( R\cup I \right) =\frac{c}{6}\ln \frac{d\left( a_+,a_- \right) d\left( b_+,b_- \right) d\left( a_+,b_+ \right) d\left( a_-,b_- \right)}{d\left( a_+,b_- \right) d\left( a_-,b_+ \right)}\;,
    \label{the von Neumann entropy of the matter fields in the region R cup I}
\end{align}
where $a_+$ and $a_-$ are the boundaries of the island region $I$, and $b_+$ and $b_-$ represent the boundaries of the radiation regions $R_+$ and $R_-$, respectively.

For the area term in the generalized entropy, we consider the cutoff surface $r = r_a$ at constant coordinate time $t$, where the metric can be described by:
\begin{align}
    \mathrm{d}{s_2}^2=\Sigma \mathrm{d}\theta ^2+\frac{\left( r^2+a^2 \right) ^2-\Delta a^2\sin ^2\theta}{\Sigma}\sin ^2\theta \mathrm{d}\phi ^2
\end{align}
The determinant of the induced metric at the cutoff surface is:
\begin{align}
    \sigma = \left[ \left( {r_a}^2 + a^2 \right) - \Delta a^2 \sin^2 \theta \right] \sin^2 \theta\;.
\end{align}
Then the area of the boundary of the island at $r = r_a$ is given by:
\begin{align}
    \text{Area} \left(r = r_a \right) = 2\pi \int_0^{\pi}{\sqrt{\left( {r_a}^2 + a^2 \right)^2 - \left( r_a - r_- \right) \left( r_a - r_+ \right) a^2 \sin^2 \theta}} \sin \theta \mathrm{d} \theta\;.
    \label{eq:the area of the boundary of the island}
\end{align}

The above integral results in a rather complicated expression for the area term, which makes the analysis very difficult to proceed. Note that the location of island is unknown in advance. However, previously studies have shown that for the eternal black holes, the location of island extends a scale proportional to the Planck length beyond the horizon. In addition, we assume that the mass parameter $M$ of the black hole is much greater than its rotational parameter $a$, i.e. $a$ is a small parameter. Combining these two aspects, the second term under the square root can be safely neglected, which gives a compact expression for the area term as 
\begin{eqnarray}
     \mathrm{Area} \left( r = r_a \right) = 4\pi\left( {r_a}^2 + a^2 \right)\;.
\end{eqnarray}

By taking the island contribution, the generalized entropy is defined as:
\begin{align}
    S_{\text{gen}}=\frac{2\mathrm{Area}\left( r = r_a \right)}{4G_N}+S_{field}\left( R\cup I \right)\;.
    \label{eq:entanglement entropy of a non-extremal black hole}
\end{align}
In the following, we will extremize the generalized entropy given in Eq.(\ref{eq:entanglement entropy of a non-extremal black hole}) at different stages of evaporation. By selecting the minimal value as the location of the island, we can determine the entanglement entropies of the radiation at early times and late times. Page demonstrated that when the subsystem is much smaller than the total system, the entanglement entropy can be approximated by the thermodynamic entropy of the subsystem \cite{page1993information, page1993average}. Therefore, we expect that in the following discussions, the entanglement entropy of the radiation will initially increase linearly at early times and approach twice the Bekenstein-Hawking entropy of the non-extremal Kerr black hole at late times.

\subsubsection{Early Stages of Evaporation}\label{subsect:Early Stages}

Now we will demonstrate that there is no island appearing in the early stages of non-extremal eternal Kerr black hole evaporation, and that the entanglement entropy increases linearly with time.

Assume that the system is in a pure state at $t=0$. By using Eq.(\ref{the von Neumann entropy of the matter fields in the region R cup I}), the generalized entropy at the early stages of black hole evaporation takes the form:
\begin{align}
    S_{\text{gen}}= & \frac{2\pi \left( {r_a}^2 + a^2 \right)}{G_N}+\frac{c}{3}\ln \left\{ \frac{\cosh \left[ \kappa \left( r_{b}^{*}-r_{a}^{*} \right) \right] -\cosh \left[ \kappa \left( t_b-t_a \right) \right]}{\cosh \left[ \kappa \left( r_{b}^{*}-r_{a}^{*} \right) \right] +\cosh \left[ \kappa \left( t_b+t_a \right) \right]} \right\} \notag \\
                    & +\frac{c}{6}\ln \left[ \frac{2^4f\left( r_a \right) f\left( r_b \right) \cosh ^2\left( \kappa t_a \right) \cosh ^2\left( \kappa t_b \right)}{\kappa ^4} \right]\;.
\end{align}
At early times, where $t_a$ and $t_b$ are much less than $r_b$, the entropy can be approximated as:
\begin{align}
    S_{\text{gen}}^{\left( early \right)}\approx \frac{2\pi \left( {r_a}^2 + a^2 \right)}{G_N}+\frac{c}{6}\ln \left[ \frac{2^4f\left( r_a \right) f\left( r_b \right) \cosh ^2\left( \kappa t_a \right) \cosh ^2\left( \kappa t_b \right)}{\kappa ^4} \right]\;.
    \label{eq:the generalized entropy of the non-extremal black hole in early stage}
\end{align}
To extremize the variations of $t_a$ and $r_a$ in the above Equation (\ref{eq:the generalized entropy of the non-extremal black hole in early stage}), we have:
\begin{align}
    \frac{\partial S_{\text{gen}}^{\left( early \right)}}{\partial t_a} & =\frac{c\kappa \sinh \left( \kappa t_a \right)}{3\cosh \left( \kappa t_a \right)}=0 \;,     \\
    \frac{\partial S_{\text{gen}}^{\left( early \right)}}{\partial r_a} & =\frac{c}{6}\left[ \frac{24\pi r_a\left( r_a-r_+ \right) \left( r_a-r_- \right) +cG_N\left( 2r_a-r_+-r_- \right)}{cG_N\left( r_a-r_+ \right) \left( r_a-r_- \right)} \right] =0 \;.
\end{align}
Neglecting the higher-order term $o(cG_N)$, the positions of the island boundaries $a_+$ and $a_-$ are approximately given by:
\begin{align}
    t_a=0, \quad r_a\approx \frac{cG_N\left( r_++r_- \right)}{24\pi r_+r_-} \approx \frac{c\left( r_++r_- \right)}{24\pi r_+r_-} {\ell_p}^2=\frac{cM}{12\pi a^2}{\ell_p}^2\;,
\end{align}
where $\ell_p$ is the Planck length. We should discard all Planck scale physics.  Therefore, this indicates that no island exists in the early stage of evaporation in an eternal non-extremal Kerr black hole. 

\subsubsection{Late Stages of Evaporation}\label{subsect:Late Stages}

We now calculate the entanglement entropy of the radiation in the Kerr black hole and verify whether the configuration of the island can resolve the information paradox in the late stages of black hole evaporation. At late times, the contribution from radiation outside the cutoff will correspondingly increase as more radiation enters the cutoff surface. While the coarse-grained entropy can continue to grow linearly, a sustained linear increase in the fine-grained entropy would violate unitarity. To solve the information paradox, we must limit the growth of the entanglement entropy.

At the late stages of evaporation, we have $t_a, t_b \gg \kappa$. Since the geodesic distance between the left wedge and the right wedge in the Penrose diagram shown in Figure \ref{Penrose diagram with island} is very large, we have:
\begin{align}
    d\left( a_+,a_- \right) \approx d\left( b_+,b_- \right) \approx d\left( a_{\pm},b_{\mp} \right) \gg d\left( a_{\pm},b_{\pm} \right)\;.
\end{align}
Therefore the generalized entropy in equation (\ref{eq:entanglement entropy of a non-extremal black hole}) can be approximated as:
\begin{align}
    S_{\text{gen}}^{\left( \text{late} \right)}\approx & \frac{2\pi \left( {r_a}^2+a^2 \right)}{G_N}+\frac{c}{6}\ln \left[ \left( r_a-r_+ \right) \left( r_a-r_- \right) \left( r_b-r_+ \right) \left( r_b-r_- \right) \right] \notag                                                                 \\
 & +\frac{c}{3}\ln \left[ \frac{8\left( {r_+}^2+a^2 \right)}{\left( r_+-r_- \right) ^2} \right] +\frac{c}{3}\ln \left\{ \cosh \left[ \kappa \left( {r_a}^*-{r_b}^* \right) \right] -\cosh \left[ \kappa \left( t_a-t_b \right) \right] \right\}\;.
    \label{eq:generalized entropy of the nonextreme black hole in late time}
\end{align}
By finding the partial derivative with respect to $t_a$ of the above equation and setting it to zero, we have:
\begin{align}
    \frac{\partial S_{\text{gen}}^{\left( \text{late} \right)}}{\partial t_a}=\frac{c}{3}\cdot \frac{-\kappa \sinh \left[ \kappa \left( t_a-t_b \right) \right]}{\cosh \left[ \kappa \left( {r_a}^*-{r_b}^* \right) \right] -\cosh \left[ \kappa \left( t_a-t_b \right) \right]}=0\;.
\end{align}
Solving this equation, we find $t_a = t_b$. Substituting $t_a = t_b$ back into $S_{\text{gen}}^{\left(\text{late}\right)}$ and then finding the extremum with respect to $r_a$, we have:
\begin{align}
    \frac{\partial S_{\text{gen}}^{\left( \text{late} \right)}}{\partial r_a}= & \frac{4\pi r_a}{G_N}+\frac{c}{3}\cdot \frac{\left( r_+-r_- \right)}{\left( r_a-r_- \right)}\cdot \frac{\sinh \left[ \kappa \left( {r_a}^*-{r_b}^* \right) \right]}{\cosh \left[ \kappa \left( {r_a}^*-{r_b}^* \right) \right] -1} \notag \\
 & +\frac{c}{6}\cdot \frac{2r_a-r_+-r_-}{\left( r_a-r_+ \right) \left( r_a-r_- \right)}=0\;.
\end{align}
Solving this equation, we find $t_a = t_b$. Substituting it back into $S_{\text{gen}}^{\left(\text{late}\right)}$ and then finding the extremum with respect to $r_a$, we have:
\begin{align}
    \frac{\partial S_{\text{gen}}^{\left( \text{late} \right)}}{\partial r_a}= & \frac{4\pi r_a}{G_N}+\frac{c}{3}\cdot \frac{\left( r_+-r_- \right)}{\left( r_a-r_- \right)}\cdot \frac{\sinh \left[ \kappa \left( {r_a}^*-{r_b}^* \right) \right]}{\cosh \left[ \kappa \left( {r_a}^*-{r_b}^* \right) \right] -1} \notag \\
    & +\frac{c}{6}\cdot \frac{2r_a-r_+-r_-}{\left( r_a-r_+ \right) \left( r_a-r_- \right)}=0\;.
\end{align}
This is a rather complicated equation for $r_a$. By expanding $\partial S_{\text{gen}}^{\left(\text{late}\right)}/\partial r_a$ at $r_+$ and neglecting higher-order terms, we have:
\begin{align}
    \frac{\partial S_{\text{gen}}^{\left( \text{late} \right)}}{\partial r_a}\approx -\frac{c}{3}\sqrt{\frac{\left( r_b-r_- \right)}{\left( r_b-r_+ \right) \left( r_+-r_- \right)}}\cdot \frac{1}{\left( r_a-r_+ \right) ^{\frac{1}{2}}}+\frac{12\pi r_+r_b-cG_N-12{\pi r_+}^2}{3G_N\left( r_b-r_+ \right)}=0\;.
\end{align}
Solving this equation, we find the location of the island as 
\begin{align}
    r_a\approx r_++\frac{c^2{G_N}^2\left( r_b-r_+ \right) \left( r_b-r_- \right)}{\left( r_+-r_- \right) \left[ 12\pi r_+\left( r_b-r_+ \right) -cG_N \right] ^2}\;.
\end{align}
It can seen that the location of the island extends a scale proportional to the Planck length beyond the event horizon as expected.
In addition, increasing the rotation parameter $a$ reduces the leading term of $r_a$, while it increases the subleading term. 

Since $r_a - r_+$ is a rather small quantity and $r_a^\ast \rightarrow -\infty$ when $r \rightarrow (r_+)_+$, we have the following approximation:
\begin{align}
    \cosh \left[ \kappa \left( r_{a}^{*}-r_{b}^{*} \right) \right] -1\approx \frac{e^{-\kappa r_{a}^{*}}e^{\kappa r_{b}^{*}}}{2}-1\approx \frac{\left( r_+-r_- \right) ^2\left[ 12\pi r_+\left( r_b-r_+ \right) -cG_N \right] ^2}{2c^2{G_N}^2\left( r_b-r_- \right) ^2}\;.
\end{align}
Substituting these approximations back into Equation (\ref{eq:generalized entropy of the nonextreme black hole in late time}), the generalized entropy at the late stages of evaporation is:
\begin{align}
    S_{\text{gen}}^{\left( \text{late} \right)}= & \frac{c}{3}\ln \left\{ \frac{4\left( {r_+}^2+a^2 \right) \left( r_b-r_+ \right) \left[ 12\pi r_+\left( r_b-r_+ \right) -cG_N \right]}{cG_N\left( r_b-r_+ \right)} \right\} \notag \\
& +\frac{2\pi \left( {r_+}^2+a^2 \right)}{G_N}\approx 2S_{\text{BH}}\;.
\end{align}

In the above expression for the entanglement entropy of the radiation, the leading term $2S_{\text{BH}}$ comes from the area of the boundary of the island. The subleading term originates from the quantum effects of the matter field, which is negligible in comparison with the leading term. Thus, in the late stages of black hole evaporation, the entanglement entropy of the Hawking radiation will approach a constant value of twice of the Bekenstein-Hawking entropy of the eternal Kerr black hole. The result also demonstrates that the single-island assumption is reasonable.

\begin{figure}
    \centering
    \includegraphics[width=8cm]{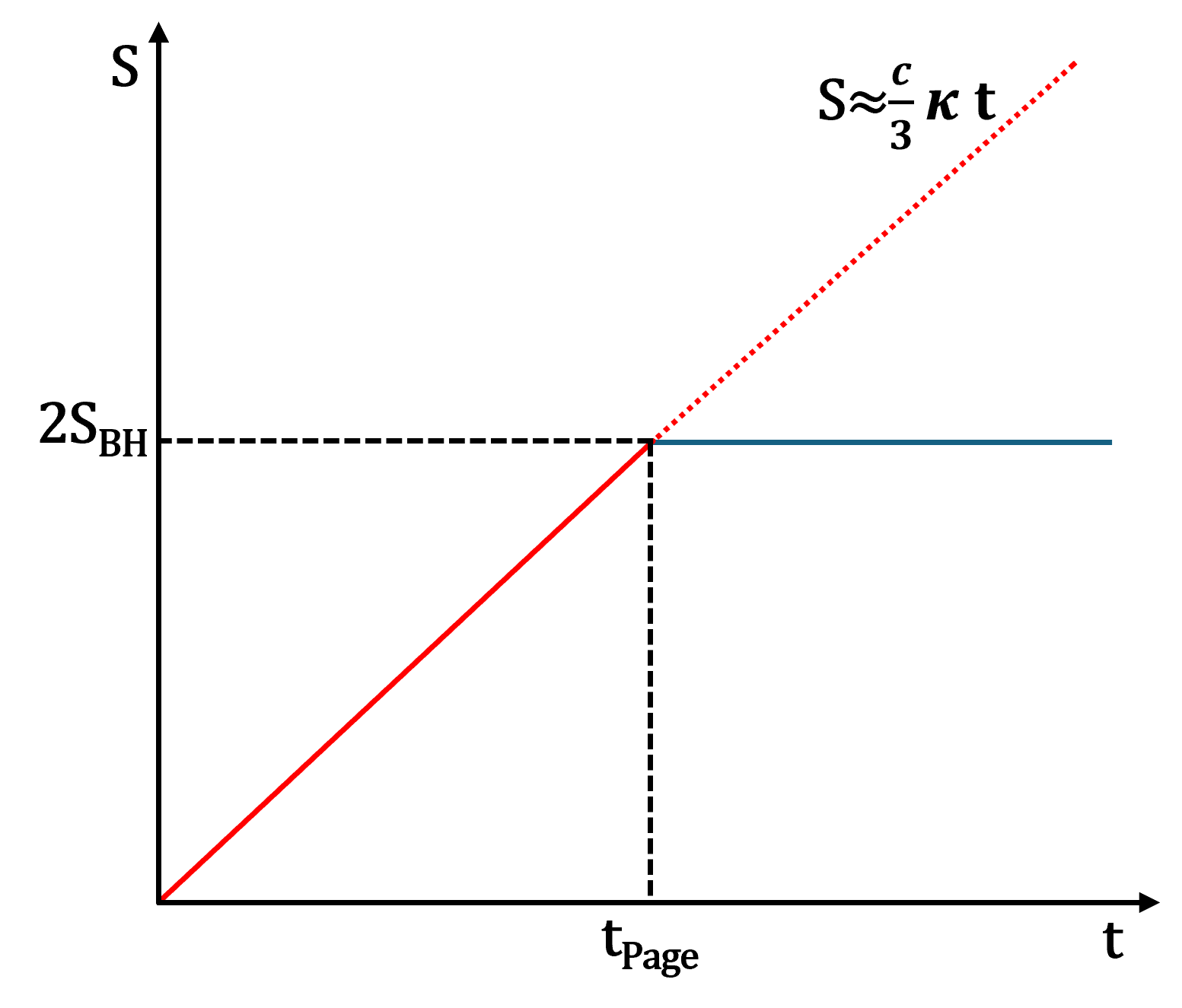}
    \caption{Evolution diagram of the entanglement entropy of Hawking radiation emitted by a non-extremal Kerr black hole. The solid red line and the dashed red line represent results without the island, while the dark blue line represents the thermodynamic entropy of the black hole. The solid red line and the dark blue line together constitute the Page curve of the eternal non-extremal Kerr black hole.}
    \label{Page curve for a non-extremal black hole}
\end{figure}

Along with the previously obtained results, we can draw the Page curve for a non-extremal Kerr black hole, as shown in Figure \ref{Page curve for a non-extremal black hole}. During the early stage, the generalized entropy increases almost linearly, while during the late stage, the increase in entanglement entropy ceases. Finally, the generalized entropy of the Hawking radiation will asymptotically approach twice the Bekenstein-Hawking entropy. Therefore, the generalized entropy of Hawking radiation from the non-extremal Kerr black hole follows the Page curve, which implies that the island rule can be properly applied to resolve the information paradox of the non-extremal Kerr black holes.

\subsection{Page time and Scrambling time}\label{subsect:Page time and Scrambling time}

In this subsection, we will discuss the Page time and scrambling time. More precisely, the Page time is when the entanglement entropy in Hawking radiation is maximal. For an evaporating black hole, the Page time is when the entanglement entropy begins to decrease. In the case of eternal black holes, the entanglement entropy of the Hawking radiation remains approximately constant after the Page time. At that point, we can determine the Page time by comparing the no-island and island entropy stages. The two curves intersect approximately at the point where a transition from a no-island state to an island state occurs, and the entanglement entropy becomes stable. Thus, the Page time can be calculated as:
\begin{align}
    t_{\text{Page}}=\frac{6S_{\text{BH}}}{c\kappa}=\frac{3\beta S_{\text{BH}}}{c \pi}\;,
\end{align}
It is obvious that the Page time decreases as the rotation parameter $a$ increases.

According to the Hayden-Preskill protocol \cite{Hayden:2007cs}, the scrambling time determines how long it takes for information that falls into the black hole to be decoded from the outgoing Hawking radiation. The information eventually enters the island region, and the scrambling time is also the time it takes for information to enter the island. Since the island region is part of the radiation region, once the lost information enters the island region, it purifies the radiation entropy.

The escaped information is stored in the Hawking radiation, and after the scrambling time, an external observer can retrieve the information that fell into the black hole from the radiation. We assume that the information falling into the black hole can be immediately decoded. The relationship between the position of the island boundary and the scrambling time is as follows: a light ray emitted from the cutoff surface intersects the boundary of the island at $r = r_a$ after one scrambling time. 

In the case of an external non-extremal Kerr black hole, the position of the island boundary is fixed near the outer horizon. Therefore, the scrambling time is actually the time it takes for a light ray released from the cutoff surface to travel to the island. For example, assume an observer at $r = r_b$ sends out a light signal at time $t_1$ and reaches the island boundary at $r = r_a$ at time $t_2$. The distance between these two points in the ingoing null direction is given by:
\begin{align}
    V\left( t_1,b \right) -V\left( t_2,a \right) =\left( t_1+{r_b}^* \right) -\left( t_2+{r_a}^* \right)\;.
\end{align}
Then the time required for information sent from the cutoff surface at $r = r_b$ towards the Kerr black hole to reach the island boundary at $r = r_a$ will be:
\begin{align}
    \Delta t=\left( {r_b}^*-{r_a}^* \right) -\left[ V\left( t_1,r_b \right) -V\left( t_2,r_a \right) \right]\;,
\end{align}
where $ V\left(t_2,r_a\right) $ should be greater than or equal to $ V\left(t_1,r_b\right) $. The minimum time required for the information to be retrieved from the Kerr black hole is given by:
\begin{align}
    t_{\text{scr}}=\Delta t={r_b}^*-{r_a}^*\;.
\end{align}
Incorporating the coordinates of the island boundary, the expression for the scrambling time is given by
\begin{align}
    t_{\text{scr}} & =\frac{1}{\kappa}\ln \left\{ \frac{\left( r_+-r_- \right) \left[ 12\pi r_+\left( r_b-r_- \right) -cG_N \right]}{cG_N\left( r_b-r_- \right)} \right\}  \notag         \\
& =\frac{\beta}{2\pi}\ln \left( S_{\text{BH}} \right) +\frac{\beta}{2\pi}\ln \left\{ \frac{\left( r_+-r_- \right) \left[ 12\pi r_+\left( r_b-r_- \right) -cG_N \right]}{\pi c\left( r_b-r_- \right) \left( {r_+}^2+a^2 \right)} \right\} \notag \\
 & \approx \frac{\beta}{2\pi}\ln \left( S_{\text{BH}} \right) +\frac{\beta}{2\pi}\ln \left\{ \frac{12\left( {r_+}^2-a^2 \right)}{c\left( {r_+}^2+a^2 \right)} \right\}\;.
\end{align}
It can be seen that increasing the rotation parameter $a$ reduces the leading term of $r_a$, while it increases the subleading term. However, assuming that the black hole's rotation parameter $a$ is much smaller than its mass $M$ and the central charge $c$ is much larger than $1$ but much smaller than its mass $M$, the last term can be neglected. Then, one can find that the scrambling time is logarithmically smaller than the lifetime of the black hole and is negligible compared to the Page time $t_{\text{Page}} = {3\beta S_{\text{BH}}}/{\pi}$. It agrees with the result predicted by the Hayden-Preskill protocol \cite{Hayden:2007cs, sekino2008fast}. This verifies the conclusion that if a non-extremal Kerr black hole acts as a quantum computer, it can perform fast-scrambling.

\section{Extremal Kerr Black Holes}
\label{sec:extremal_BHs}
\renewcommand{\theequation}{4.\arabic{equation}}
\setcounter{equation}{0}

Since the extremal Kerr black hole has different Penrose diagram compared to the non-extremal Kerr black hole, the entanglement entropy of Hawking radiation for extremal Kerr black holes cannot obtained by  straightforwardly taking the limit $r_- \rightarrow r_+$ for non-extremal Kerr black holes. In this chapter, we will recalculate the entanglement entropy of Hawking radiation for extremal Kerr black holes using a method similar to that used for non-extremal Kerr black holes.

\begin{figure}
    \centering
    \includegraphics[width=8cm]{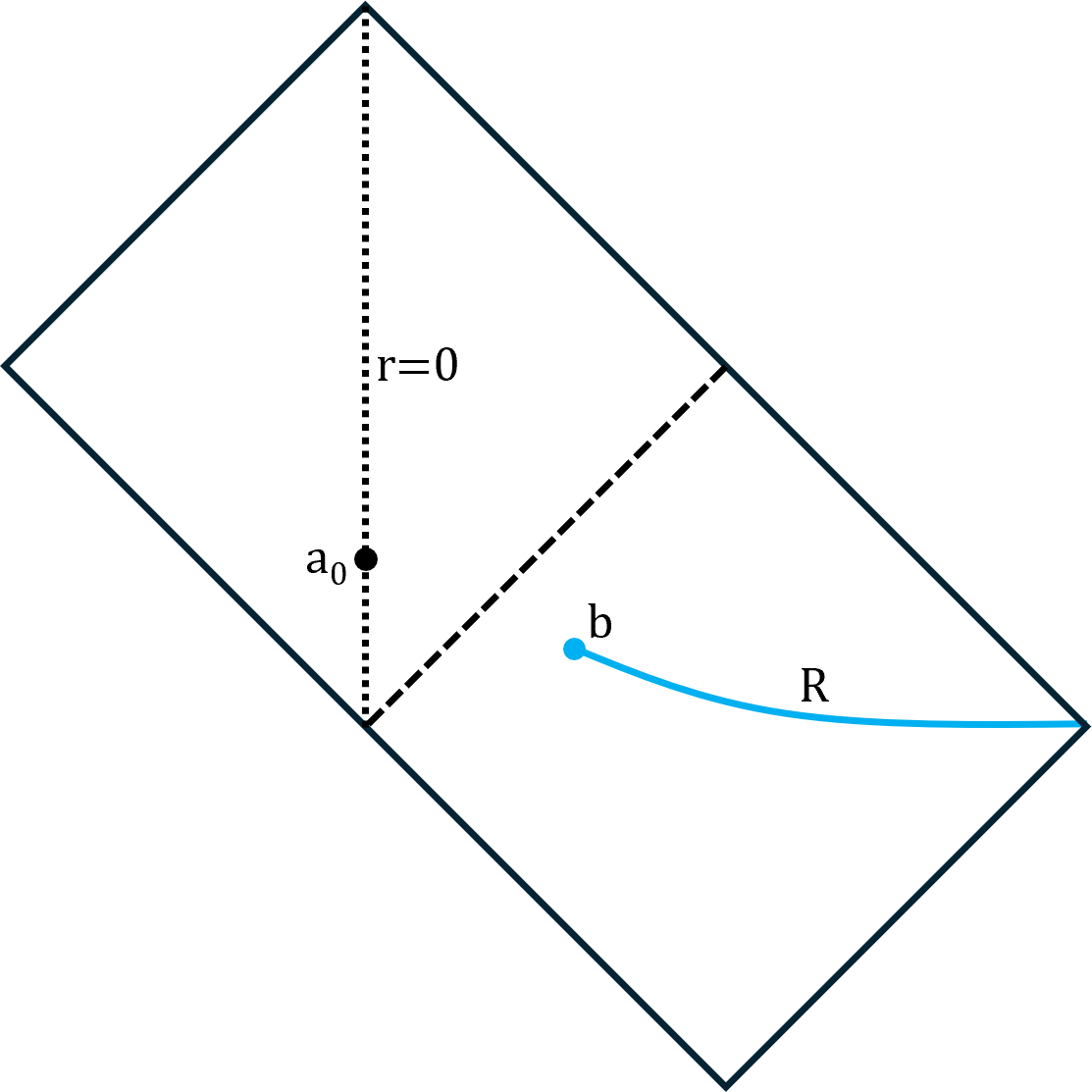}\\
    \caption{Penrose diagram of an eternal extremal Kerr black hole without island. The Hawking radiation is assumed to be located in the regions $R$, with boundaries $b$.}
    \label{Penrose diagram without island in extreme case}
\end{figure}

As shown in Figure \ref{Penrose diagram without island in extreme case}, the Hawking radiation is collected in the region $R$. We assume that this region is far from the event horizon of the black hole, allowing the dynamics of the Hawking radiation to be approximated by a conformal field theory in flat spacetime. We assume that the entire system is in a pure state at $t=0$. In the early stages of evaporation of extremal Kerr black holes, the radiation region extends from the inner boundary $b$ to the spacial infinity, which makes the geodesic length for the radiation divergent. Therefore, this approach is ineffective. However, from the physical consideration, we believe that the entanglement entropy of the radiation at early time is linearly increasing with the time.

\begin{figure}
    \centering
    \includegraphics[width=8cm]{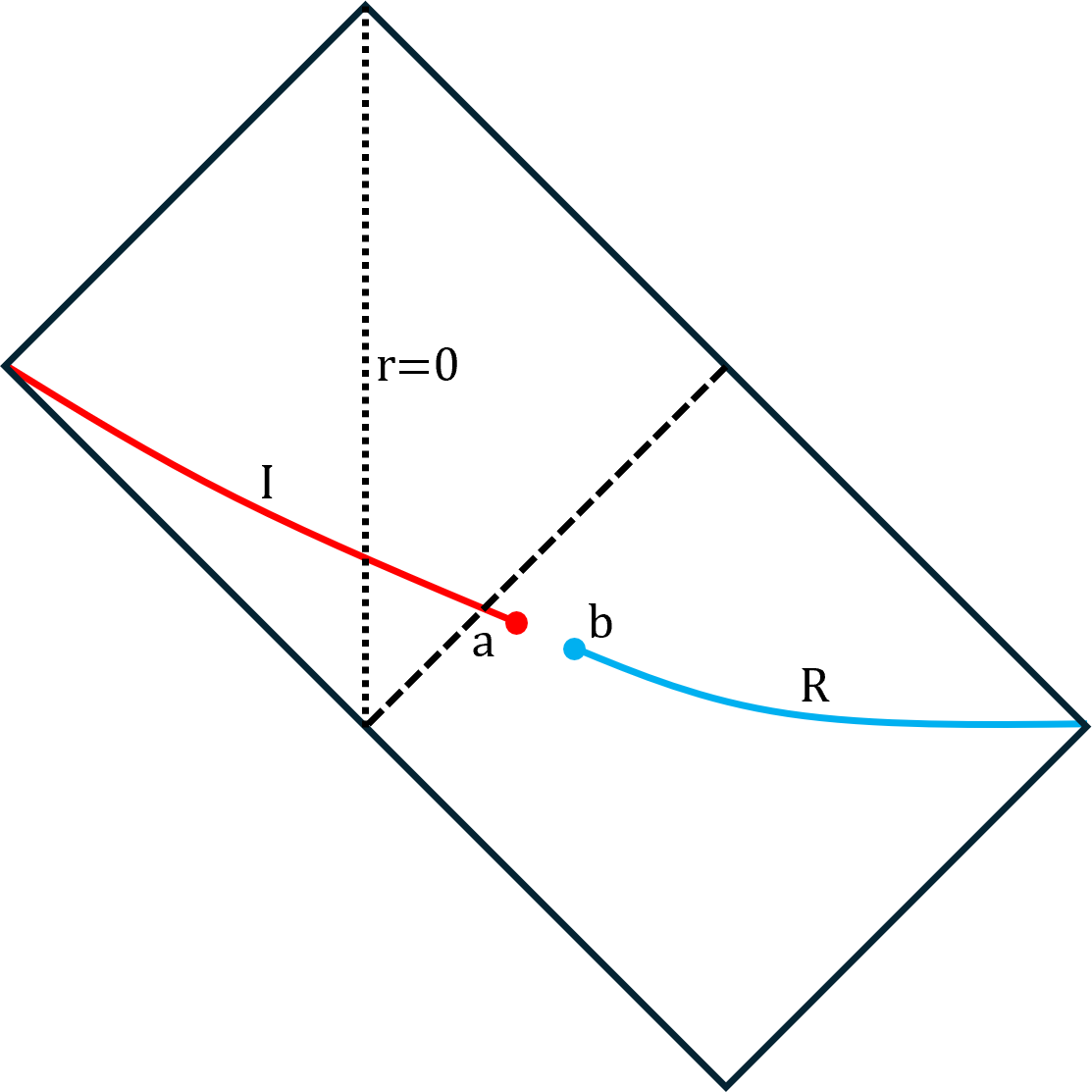}
    \caption{Penrose diagram of an eternal extremal Kerr black hole with the island. The Hawking radiation is assumed to be located in the regions $R$, with boundary $b$. The island region, denoted as $I$, has boundary $a$.}
    \label{Penrose diagram with island in extreme case}
\end{figure}

We now consider the question that whether one can get a finite entanglement entropy of the radiation for the extremal Kerr black hole at late times. It is expected that at late times, the entanglement island emerges. As shown in Figure \ref{Penrose diagram with island in extreme case}, we introduce an island region $I$ with its boundary denoted as $a$. In this diagram the singularity ring $r=0$ can be safely crossed and the $r<0$ region is also meaningful. The Cauchy surface must intersect the surface $r=0$. It is assumed that the intersection point is at $a_0=\left(t_0,0\right)$. To determine the entanglement entropy of the Hawking radiation, the geodesic we seek must include the point $a_0$. However, because the extremal Kerr spacetime has a singularity at $a_0$, the expression for the geodesic distance is not well-defined. This makes calculating the entanglement entropy of Hawking radiation in the extremal case quite challenging. In \cite{Ahn:2021chg}, a similar problem is encountered when studying the extremal dilaton black holes.

One key observation is that the complementary region $[a, b]$ of the radiation and the island does not contain any singularity. Thus, in the late stages of extremal Kerr black hole evaporation, the geodesic length between $a$ and $b$ is well-defined and the generalized entropy can be calculated from this geodesic length. In this sense, it is possible to find the island's position and study the behavior of entanglement entropy at late times by extremalizing the generalized entropy.

In the late stages of evaporation, we should have $r_a \approx r_h$. Using the same approximation as in Equation (\ref{eq:the area of the boundary of the island}), we obtain the generalized entropy of Hawking radiation for the extremal case as:
\begin{align}
    S_{\text{gen}}^{\left( \text{late} \right)}&=\frac{Area\left( \partial I \right)}{4G_N}+\frac{c}{6}\ln \left( d\left( a,b \right) \right) \notag \\
    &\approx \frac{\pi \left( {r_a}^2+a^2 \right)}{G_N}+\frac{c}{6}\left\{ \frac{2{r_h}^2\left( r_a-r_h \right) \left( r_b-r_h \right) \left[ \cosh \left( \frac{r_{a}^{*}-r_{b}^{*}}{r_h} \right) -\cosh \left( \frac{\xi _a-\xi _b}{r_h} \right) \right]}{{r_h}^2+a^2} \right\}\;.
    \label{eq:the generalized entropy of Hawking radiation for extreme case}
\end{align}
By finding the partial derivative with respect to $t_a$ of the above equation and setting it to zero, we have:
\begin{align}
    \frac{\partial S_{\text{gen}}^{\left( \text{late} \right)}}{\partial t_a}=\frac{c}{6}\frac{-\sinh \left( \frac{t_a-t_b}{r_h} \right)}{\cosh \left( \frac{r_{a}^{*}-r_{b}^{*}}{r_h} \right) -\cosh \left( \frac{t_a-t_b}{r_h} \right)}=0\;.
\end{align}
Solving this equation, we find $t_a = t_b$. Substituting it back into $S_{\text{gen}}^{\left(\text{late}\right)}$ and taking the partial derivative with respect to $r_a$, we have:
\begin{align}
    \frac{\partial S_{\text{gen}}^{\left( \text{late} \right)}}{\partial r_a}=\frac{2\pi r_a}{G_N}+\frac{c}{6}\left\{ \frac{2r_h\sinh \left( \frac{r_{a}^{*}-r_{b}^{*}}{r_h} \right)}{\left( r_a-r_h \right) ^2\left[ \cosh \left( \frac{r_{a}^{*}-r_{b}^{*}}{r_h} \right) -1 \right]}+\frac{1}{r_a-r_h} \right\} =0\;.
\end{align}
Setting $r_a = r_h + \alpha \sqrt{cG_N}$, and then substituting and neglecting higher-order terms of $cG_N$, we have:
\begin{align}
    r_a\approx r_h+\sqrt{\frac{cG_N}{6\pi}}\;.
\end{align}
Using the approximation in Equation (\ref{eq:the generalized entropy of Hawking radiation for extreme case}), the entanglement entropy in the late stages of extremal Kerr black hole evaporation is given by:
\begin{align}
    S_{\text{gen}}^{\left( \text{late} \right)}\approx \frac{\pi \left( {r_h}^2+a^2 \right)}{G_N}+\sqrt{\frac{2{cr_h}^2}{3G_N}}-\frac{cr_h}{3\left( r_b-r_h \right)}+\frac{c}{12}\ln \left[ \frac{cG_N\left( r_b-r_h \right) ^2}{6\pi} \right] \approx S_{\text{BH}}\;.
\end{align}
The leading term, which comes from the area of the island, is the Bekenstein-Hawking entropy of the eternal extremal Kerr black hole. The subleading term arises from the quantum effects of the matter field and is negligible compared to the leading term. Therefore, in the late stages of black hole evaporation, the entanglement entropy of Hawking radiation is finite and approach the Bekenstein-Hawking entropy. 

\section{Conclusion and discussion}
\label{sec:con_disc}
\renewcommand{\theequation}{7.\arabic{equation}}
\setcounter{equation}{0}

In this paper, we investigated the black hole information problem in four-dimensional Kerr spacetime. We first reduced the four-dimensional Kerr metric to a two-dimensional form and then calculated the entanglement entropy of the radiation region within this two-dimensional framework. In gravitational systems, the quantum entanglement entropy is the minimum of all extremal values of the sum of the island's area entropy and the von Neumann entropy of the radiation in the entanglement region. We began by studying non-extremal Kerr black holes. In the early stages of evaporation, no island form because insufficient radiation is produced initially. Consequently, the contribution to the entanglement entropy mainly comes from the radiation itself, and no island exist. In the late stages of black hole evaporation, radiation becomes the dominant term, and the radiation region primarily entangled with the island region. By introducing an island region within the Kerr spacetime, we obtained a scrambling time following the Hayden-Preskill protocol, consistent with the Page time. We further studied extremal Kerr black holes. In the early stages of black hole evaporation, calculating the von Neumann entropy of the matter fields is challenging since the island is located at $r<0$. However, in the late stages, we showed that the entanglement entropy of Hawking radiation will approach the Bekenstein-Hawking entropy.

For future directions, it is meaningful to generalize the present work to the evaporating Kerr black hole. The Page curve for the evaporating Kerr black hole was previously studied in \cite{Nian:2019buz}. Note that there is a possible resolution to the information paradox for the evaporating Kerr black hole by treating it as a quantum wormhole \cite{Nian:2023xmr}. It is also interesting to extend the discussion to other types of rotating black holes. This study has examined four-dimensional eternal Kerr black hole in asymptotically flat spacetime. In asymptotically flat spacetime, there is no need to couple the system with a bath to collect the radiation. We plan to address the information loss problem of the four-dimensional Kerr-AdS black hole using similar methods as presented here. In addition, further investigation into multi-island scenarios might provide a more detailed description of the Page curve at the Page time.

\bibliographystyle{unsrt}
\bibliography{reference}

\begin{thebibliography}{100}

\bibitem{hawking1975particle}
Stephen~W Hawking.
\newblock Particle creation by black holes.
\newblock {\em Communications in mathematical physics}, 43(3):199--220, 1975.

\bibitem{Wald:1975kc}
Robert~M. Wald.
\newblock {On Particle Creation by Black Holes}.
\newblock {\em Commun. Math. Phys.}, 45:9--34, 1975.

\bibitem{Unruh:1976db}
W.~G. Unruh.
\newblock {Notes on black hole evaporation}.
\newblock {\em Phys. Rev. D}, 14:870, 1976.

\bibitem{hawking1976breakdown}
Stephen~W Hawking.
\newblock Breakdown of predictability in gravitational collapse.
\newblock {\em Physical Review D}, 14(10):2460, 1976.

\bibitem{Almheiri:2020cfm}
Ahmed Almheiri, Thomas Hartman, Juan Maldacena, Edgar Shaghoulian, and Amirhossein Tajdini.
\newblock {The entropy of Hawking radiation}.
\newblock {\em Rev. Mod. Phys.}, 93(3):035002, 2021.

\bibitem{page1993information}
Don~N Page.
\newblock Information in black hole radiation.
\newblock {\em Physical review letters}, 71(23):3743, 1993.

\bibitem{page2013time}
Don~N Page.
\newblock Time dependence of hawking radiation entropy.
\newblock {\em Journal of Cosmology and Astroparticle Physics}, 2013(09):028, 2013.

\bibitem{Page:1993df}
Don~N. Page.
\newblock {Average entropy of a subsystem}.
\newblock {\em Phys. Rev. Lett.}, 71:1291--1294, 1993.

\bibitem{harlow2016jerusalem}
Daniel Harlow.
\newblock Jerusalem lectures on black holes and quantum information.
\newblock {\em Reviews of Modern Physics}, 88(1):015002, 2016.

\bibitem{maldacena1999large}
Juan Maldacena.
\newblock The large-n limit of superconformal field theories and supergravity.
\newblock {\em International journal of theoretical physics}, 38(4):1113--1133, 1999.

\bibitem{almheiri2013black}
Ahmed Almheiri, Donald Marolf, Joseph Polchinski, and James Sully.
\newblock Black holes: complementarity or firewalls?
\newblock {\em Journal of High Energy Physics}, 2013(2):1--20, 2013.

\bibitem{Penington:2019npb}
Geoffrey Penington.
\newblock {Entanglement Wedge Reconstruction and the Information Paradox}.
\newblock {\em JHEP}, 09:002, 2020.

\bibitem{Almheiri:2019psf}
Ahmed Almheiri, Netta Engelhardt, Donald Marolf, and Henry Maxfield.
\newblock {The entropy of bulk quantum fields and the entanglement wedge of an evaporating black hole}.
\newblock {\em JHEP}, 12:063, 2019.

\bibitem{Almheiri:2019hni}
Ahmed Almheiri, Raghu Mahajan, Juan Maldacena, and Ying Zhao.
\newblock {The Page curve of Hawking radiation from semiclassical geometry}.
\newblock {\em JHEP}, 03:149, 2020.

\bibitem{Penington:2019kki}
Geoff Penington, Stephen~H. Shenker, Douglas Stanford, and Zhenbin Yang.
\newblock {Replica wormholes and the black hole interior}.
\newblock {\em JHEP}, 03:205, 2022.

\bibitem{Almheiri:2019qdq}
Ahmed Almheiri, Thomas Hartman, Juan Maldacena, Edgar Shaghoulian, and Amirhossein Tajdini.
\newblock {Replica Wormholes and the Entropy of Hawking Radiation}.
\newblock {\em JHEP}, 05:013, 2020.

\bibitem{ryu2006holographic}
Shinsei Ryu and Tadashi Takayanagi.
\newblock Holographic derivation of entanglement entropy from the anti-de sitter space/conformal field theory correspondence.
\newblock {\em Physical review letters}, 96(18):181602, 2006.

\bibitem{hubeny2007covariant}
Veronika~E Hubeny, Mukund Rangamani, and Tadashi Takayanagi.
\newblock A covariant holographic entanglement entropy proposal.
\newblock {\em Journal of High Energy Physics}, 2007(07):062, 2007.

\bibitem{lewkowycz2013generalized}
Aitor Lewkowycz and Juan Maldacena.
\newblock Generalized gravitational entropy.
\newblock {\em Journal of High Energy Physics}, 2013(8):1--29, 2013.

\bibitem{barrella2013holographic}
Taylor Barrella, Xi~Dong, Sean~A Hartnoll, and Victoria~L Martin.
\newblock Holographic entanglement beyond classical gravity.
\newblock {\em Journal of High Energy Physics}, 2013(9):1--36, 2013.

\bibitem{faulkner2013quantum}
Thomas Faulkner, Aitor Lewkowycz, and Juan Maldacena.
\newblock Quantum corrections to holographic entanglement entropy.
\newblock {\em Journal of High Energy Physics}, 2013(11):1--18, 2013.

\bibitem{engelhardt2015quantum}
Netta Engelhardt and Aron~C Wall.
\newblock Quantum extremal surfaces: holographic entanglement entropy beyond the classical regime.
\newblock {\em Journal of High Energy Physics}, 2015(1):1--27, 2015.

\bibitem{Hartman:2020swn}
Thomas Hartman, Edgar Shaghoulian, and Andrew Strominger.
\newblock {Islands in Asymptotically Flat 2D Gravity}.
\newblock {\em JHEP}, 07:022, 2020.

\bibitem{Goto:2020wnk}
Kanato Goto, Thomas Hartman, and Amirhossein Tajdini.
\newblock {Replica wormholes for an evaporating 2D black hole}.
\newblock {\em JHEP}, 04:289, 2021.

\bibitem{Geng:2024xpj}
Hao Geng.
\newblock {Replica Wormholes and Entanglement Islands in the Karch-Randall Braneworld}, arXiv:2405.14872.

\bibitem{Almheiri:2019yqk}
Ahmed Almheiri, Raghu Mahajan, and Juan Maldacena.
\newblock {Islands outside the horizon}.
\newblock arXiv:1910.11077.

\bibitem{Chen:2019uhq}
Hong~Zhe Chen, Zachary Fisher, Juan Hernandez, Robert~C. Myers, and Shan-Ming Ruan.
\newblock {Information Flow in Black Hole Evaporation}.
\newblock {\em JHEP}, 03:152, 2020.

\bibitem{Almheiri:2019psy}
Ahmed Almheiri, Raghu Mahajan, and Jorge~E. Santos.
\newblock {Entanglement islands in higher dimensions}.
\newblock {\em SciPost Phys.}, 9(1):001, 2020.

\bibitem{Gautason:2020tmk}
Fri\dh{}rik~Freyr Gautason, Lukas Schneiderbauer, Watse Sybesma, and L\'arus Thorlacius.
\newblock {Page Curve for an Evaporating Black Hole}.
\newblock {\em JHEP}, 05:091, 2020.

\bibitem{Anegawa:2020ezn}
Takanori Anegawa and Norihiro Iizuka.
\newblock {Notes on islands in asymptotically flat 2d dilaton black holes}.
\newblock {\em JHEP}, 07:036, 2020.

\bibitem{Hashimoto:2020cas}
Koji Hashimoto, Norihiro Iizuka, and Yoshinori Matsuo.
\newblock {Islands in Schwarzschild black holes}.
\newblock {\em JHEP}, 06:085, 2020.

\bibitem{Hollowood:2020cou}
Timothy~J. Hollowood and S.~Prem Kumar.
\newblock {Islands and Page Curves for Evaporating Black Holes in JT Gravity}.
\newblock {\em JHEP}, 08:094, 2020.

\bibitem{Krishnan:2020oun}
Chethan Krishnan, Vaishnavi Patil, and Jude Pereira.
\newblock {Page Curve and the Information Paradox in Flat Space}.
\newblock arXiv:2005.02993.

\bibitem{Alishahiha:2020qza}
Mohsen Alishahiha, Amin Faraji~Astaneh, and Ali Naseh.
\newblock {Island in the presence of higher derivative terms}.
\newblock {\em JHEP}, 02:035, 2021.

\bibitem{Chen:2020uac}
Hong~Zhe Chen, Robert~C. Myers, Dominik Neuenfeld, Ignacio~A. Reyes, and Joshua Sandor.
\newblock {Quantum Extremal Islands Made Easy, Part I: Entanglement on the Brane}.
\newblock {\em JHEP}, 10:166, 2020.

\bibitem{Geng:2020qvw}
Hao Geng and Andreas Karch.
\newblock {Massive islands}.
\newblock {\em JHEP}, 09:121, 2020.

\bibitem{Li:2020ceg}
Tianyi Li, Jinwei Chu, and Yang Zhou.
\newblock {Reflected Entropy for an Evaporating Black Hole}.
\newblock {\em JHEP}, 11:155, 2020.

\bibitem{Chandrasekaran:2020qtn}
Venkatesa Chandrasekaran, Masamichi Miyaji, and Pratik Rath.
\newblock {Including contributions from entanglement islands to the reflected entropy}.
\newblock {\em Phys. Rev. D}, 102(8):086009, 2020.

\bibitem{Bak:2020enw}
Dongsu Bak, Chanju Kim, Sang-Heon Yi, and Junggi Yoon.
\newblock {Unitarity of entanglement and islands in two-sided Janus black holes}.
\newblock {\em JHEP}, 01:155, 2021.

\bibitem{Krishnan:2020fer}
Chethan Krishnan.
\newblock {Critical Islands}.
\newblock {\em JHEP}, 01:179, 2021.

\bibitem{Chen:2020jvn}
Hong~Zhe Chen, Zachary Fisher, Juan Hernandez, Robert~C. Myers, and Shan-Ming Ruan.
\newblock {Evaporating Black Holes Coupled to a Thermal Bath}.
\newblock {\em JHEP}, 01:065, 2021.

\bibitem{Hartman:2020khs}
Thomas Hartman, Yikun Jiang, and Edgar Shaghoulian.
\newblock {Islands in cosmology}.
\newblock {\em JHEP}, 11:111, 2020.

\bibitem{Balasubramanian:2020xqf}
Vijay Balasubramanian, Arjun Kar, and Tomonori Ugajin.
\newblock {Islands in de Sitter space}.
\newblock {\em JHEP}, 02:072, 2021.

\bibitem{Balasubramanian:2020coy}
Vijay Balasubramanian, Arjun Kar, and Tomonori Ugajin.
\newblock {Entanglement between two disjoint universes}.
\newblock {\em JHEP}, 02:136, 2021.

\bibitem{Sybesma:2020fxg}
Watse Sybesma.
\newblock {Pure de Sitter space and the island moving back in time}.
\newblock {\em Class. Quant. Grav.}, 38(14):145012, 2021.

\bibitem{Chen:2020hmv}
Hong~Zhe Chen, Robert~C. Myers, Dominik Neuenfeld, Ignacio~A. Reyes, and Joshua Sandor.
\newblock {Quantum Extremal Islands Made Easy, Part II: Black Holes on the Brane}.
\newblock {\em JHEP}, 12:025, 2020.

\bibitem{Ling:2020laa}
Yi~Ling, Yuxuan Liu, and Zhuo-Yu Xian.
\newblock {Island in Charged Black Holes}.
\newblock {\em JHEP}, 03:251, 2021.

\bibitem{Hernandez:2020nem}
Juan Hernandez, Robert~C. Myers, and Shan-Ming Ruan.
\newblock {Quantum extremal islands made easy. Part III. Complexity on the brane}.
\newblock {\em JHEP}, 02:173, 2021.

\bibitem{Matsuo:2020ypv}
Yoshinori Matsuo.
\newblock {Islands and stretched horizon}.
\newblock {\em JHEP}, 07:051, 2021.

\bibitem{Karananas:2020fwx}
Georgios~K. Karananas, Alex Kehagias, and John Taskas.
\newblock {Islands in linear dilaton black holes}.
\newblock {\em JHEP}, 03:253, 2021.

\bibitem{Wang:2021woy}
Xuanhua Wang, Ran Li, and Jin Wang.
\newblock {Islands and Page curves of Reissner-Nordstr\"om black holes}.
\newblock {\em JHEP}, 04:103, 2021.

\bibitem{Geng:2021wcq}
Hao Geng, Yasunori Nomura, and Hao-Yu Sun.
\newblock {Information paradox and its resolution in de Sitter holography}.
\newblock {\em Phys. Rev. D}, 103(12):126004, 2021.

\bibitem{Fallows:2021sge}
Seamus Fallows and Simon~F. Ross.
\newblock {Islands and mixed states in closed universes}.
\newblock {\em JHEP}, 07:022, 2021.

\bibitem{Bhattacharya:2021jrn}
Aranya Bhattacharya, Arpan Bhattacharyya, Pratik Nandy, and Ayan~K. Patra.
\newblock {Islands and complexity of eternal black hole and radiation subsystems for a doubly holographic model}.
\newblock {\em JHEP}, 05:135, 2021.

\bibitem{Kim:2021gzd}
Wontae Kim and Mungon Nam.
\newblock {Entanglement entropy of asymptotically flat non-extremal and extremal black holes with an island}.
\newblock {\em Eur. Phys. J. C}, 81(10):869, 2021.

\bibitem{Wang:2021mqq}
Xuanhua Wang, Ran Li, and Jin Wang.
\newblock {Page curves for a family of exactly solvable evaporating black holes}.
\newblock {\em Phys. Rev. D}, 103(12):126026, 2021.

\bibitem{Geng:2021iyq}
Hao Geng, Severin L\"ust, Rashmish~K. Mishra, and David Wakeham.
\newblock {Holographic BCFTs and Communicating Black Holes}.
\newblock {\em jhep}, 08:003, 2021.

\bibitem{Uhlemann:2021nhu}
Christoph~F. Uhlemann.
\newblock {Islands and Page curves in 4d from Type IIB}.
\newblock {\em JHEP}, 08:104, 2021.

\bibitem{Li:2021lfo}
Ran Li, Xuanhua Wang, and Jin Wang.
\newblock {Island may not save the information paradox of Liouville black holes}.
\newblock {\em Phys. Rev. D}, 104(10):106015, 2021.

\bibitem{Chu:2021gdb}
Jinwei Chu, Feiyu Deng, and Yang Zhou.
\newblock {Page curve from defect extremal surface and island in higher dimensions}.
\newblock {\em JHEP}, 10:149, 2021.

\bibitem{Lu:2021gmv}
Yizhou Lu and Jiong Lin.
\newblock {Islands in Kaluza\textendash{}Klein black holes}.
\newblock {\em Eur. Phys. J. C}, 82(2):132, 2022.

\bibitem{Yu:2021cgi}
Ming-Hui Yu and Xian-Hui Ge.
\newblock {Islands and Page curves in charged dilaton black holes}.
\newblock {\em Eur. Phys. J. C}, 82(1):14, 2022.

\bibitem{Ahn:2021chg}
Byoungjoon Ahn, Sang-Eon Bak, Hyun-Sik Jeong, Keun-Young Kim, and Ya-Wen Sun.
\newblock {Islands in charged linear dilaton black holes}.
\newblock {\em Phys. Rev. D}, 105(4):046012, 2022.

\bibitem{Aguilar-Gutierrez:2021bns}
Sergio~E. Aguilar-Gutierrez, Aidan Chatwin-Davies, Thomas Hertog, Natalia Pinzani-Fokeeva, and Brandon Robinson.
\newblock {Islands in Multiverse Models}.
\newblock {\em JHEP}, 11:212, 2021.
\newblock [Addendum: JHEP 05, 137 (2022), Erratum: JHEP 05, 082 (2022)].

\bibitem{Kames-King:2021etp}
Joshua Kames-King, Evita M.~H. Verheijden, and Erik~P. Verlinde.
\newblock {No Page curves for the de Sitter horizon}.
\newblock {\em JHEP}, 03:040, 2022.

\bibitem{Cao:2021ujs}
Nam~H. Cao.
\newblock {Entanglement entropy and Page curve of black holes with island in massive gravity}.
\newblock {\em Eur. Phys. J. C}, 82(4):381, 2022.

\bibitem{Saha:2021ohr}
Ashis Saha, Sunandan Gangopadhyay, and Jyoti~Prasad Saha.
\newblock {Mutual information, islands in black holes and the Page curve}.
\newblock {\em Eur. Phys. J. C}, 82(5):476, 2022.

\bibitem{Azarnia:2021uch}
Sanam Azarnia, Reza Fareghbal, Ali Naseh, and Hamed Zolfi.
\newblock {Islands in flat-space cosmology}.
\newblock {\em Phys. Rev. D}, 104(12):126017, 2021.

\bibitem{Okuyama:2021bqg}
Kazumi Okuyama and Kazuhiro Sakai.
\newblock {Page curve from dynamical branes in JT gravity}.
\newblock {\em JHEP}, 02:087, 2022.

\bibitem{Omidi:2021opl}
Farzad Omidi.
\newblock {Entropy of Hawking radiation for two-sided hyperscaling violating black branes}.
\newblock {\em JHEP}, 04:022, 2022.

\bibitem{Yu:2021rfg}
Ming-Hui Yu, Cheng-Yuan Lu, Xian-Hui Ge, and Sang-Jin Sin.
\newblock {Island, Page curve, and superradiance of rotating BTZ black holes}.
\newblock {\em Phys. Rev. D}, 105(6):066009, 2022.

\bibitem{Gan:2022jay}
Wen-Cong Gan, Dong-Hui Du, and Fu-Wen Shu.
\newblock {Island and Page curve for one-sided asymptotically flat black hole}.
\newblock {\em JHEP}, 07:020, 2022.

\bibitem{Seo:2022ezk}
Min-Seok Seo.
\newblock {Information paradox and island in quasi-de Sitter space}.
\newblock {\em Eur. Phys. J. C}, 82(12):1082, 2022.

\bibitem{Azarnia:2022kmp}
Sanam Azarnia and Reza Fareghbal.
\newblock {Islands in Kerr\textendash{}de Sitter spacetime and their flat limit}.
\newblock {\em Phys. Rev. D}, 106(2):026012, 2022.

\bibitem{Tian:2022pso}
Jia Tian.
\newblock {Islands in Generalized Dilaton Theories}.
\newblock {\em Symmetry}, 15(7):1402, 2023.

\bibitem{Afrasiar:2022ebi}
Mir Afrasiar, Jaydeep Kumar~Basak, Ashish Chandra, and Gautam Sengupta.
\newblock {Islands for entanglement negativity in communicating black holes}.
\newblock {\em Phys. Rev. D}, 108(6):066013, 2023.

\bibitem{Anand:2022mla}
Ankit Anand.
\newblock {Page curve and island in EGB gravity}.
\newblock {\em Nucl. Phys. B}, 993:116284, 2023.

\bibitem{Djordjevic:2022qdk}
Stefan Djordjevi\'c, Aleksandra Go\v{c}anin, Dragoljub Go\v{c}anin, and Voja Radovanovi\'c.
\newblock {Page curve for an eternal Schwarzschild black hole in a dimensionally reduced model of dilaton gravity}.
\newblock {\em Phys. Rev. D}, 106(10):105015, 2022.

\bibitem{Goswami:2022ylc}
Kaberi Goswami and K.~Narayan.
\newblock {Small Schwarzschild de Sitter black holes, quantum extremal surfaces and islands}.
\newblock {\em JHEP}, 10:031, 2022.

\bibitem{Yu:2022xlh}
Ming-Hui Yu and Xian-Hui Ge.
\newblock {Entanglement islands in generalized two-dimensional dilaton black holes}.
\newblock {\em Phys. Rev. D}, 107(6):066020, 2023.

\bibitem{Hu:2022zgy}
Peng-Ju Hu, Dongqi Li, and Rong-Xin Miao.
\newblock {Island on codimension-two branes in AdS/dCFT}.
\newblock {\em JHEP}, 11:008, 2022.

\bibitem{Lu:2022tmt}
Cheng-Yuan Lu, Ming-Hui Yu, Xian-Hui Ge, and Li-Jun Tian.
\newblock {Page curve and phase transition in deformed Jackiw\textendash{}Teitelboim gravity}.
\newblock {\em Eur. Phys. J. C}, 83(3):215, 2023.

\bibitem{BenDayan:2022nmb}
Ido Ben-Dayan, Merav Hadad, and Elizabeth Wildenhain.
\newblock {Islands in the fluid: islands are common in cosmology}.
\newblock {\em JHEP}, 03:077, 2023.

\bibitem{Baek:2022ozg}
Jong-Hyun Baek and Kang-Sin Choi.
\newblock {Islands in proliferating de Sitter spaces}.
\newblock {\em JHEP}, 05:098, 2023.

\bibitem{Guo:2023gfa}
Chang-Zhong Guo, Wen-Cong Gan, and Fu-Wen Shu.
\newblock {Page curves and entanglement islands for the step-function Vaidya model of evaporating black holes}.
\newblock {\em JHEP}, 05:042, 2023.

\bibitem{Parvizi:2023foz}
Shahrokh Parvizi and Mojtaba Shahbazi.
\newblock {Analogue gravity and the island prescription}.
\newblock {\em Eur. Phys. J. C}, 83(8):705, 2023.

\bibitem{Wu:2023uyb}
Chih-Hung Wu and Jiuci Xu.
\newblock {Islands in non-minimal dilaton gravity: exploring effective theories for black hole evaporation}.
\newblock {\em JHEP}, 10:094, 2023.

\bibitem{Li:2023fly}
Dongqi Li and Rong-Xin Miao.
\newblock {Massless entanglement islands in cone holography}.
\newblock {\em JHEP}, 06:056, 2023.

\bibitem{RoyChowdhury:2023eol}
Anirban Roy~Chowdhury, Ashis Saha, and Sunandan Gangopadhyay.
\newblock {Mutual information of subsystems and the Page curve for the Schwarzschild\textendash{}de Sitter black hole}.
\newblock {\em Phys. Rev. D}, 108(2):026003, 2023.

\bibitem{Wang:2023eyb}
Xuanhua Wang, Kun Zhang, and Jin Wang.
\newblock {Entanglement islands, fire walls and state paradox from quantum teleportation and entanglement swapping}.
\newblock {\em Class. Quant. Grav.}, 40(9):095012, 2023.

\bibitem{Tong:2023nvi}
Chen-Wei Tong, Dong-Hui Du, and Jia-Rui Sun.
\newblock {Island of Reissner-Nordstr\"om anti\textendash{}de Sitter black holes in the large D limit}.
\newblock {\em Phys. Rev. D}, 109(10):104053, 2024.

\bibitem{Yu:2023whl}
Ming-Hui Yu, Xian-Hui Ge, and Cheng-Yuan Lu.
\newblock {Page curves for accelerating black holes}.
\newblock {\em Eur. Phys. J. C}, 83(12):1104, 2023.

\bibitem{Chou:2023adi}
Chia-Jui Chou, Hans~B. Lao, and Yi~Yang.
\newblock {Page curve of AdS-Vaidya model for evaporating black holes}.
\newblock {\em JHEP}, 05:342, 2024.

\bibitem{Chang:2023gkt}
Jing-Cheng Chang, Song He, Yu-Xiao Liu, and Long Zhao.
\newblock {Island formula in Planck brane}.
\newblock {\em JHEP}, 11:006, 2023.

\bibitem{Matsuo:2023cmb}
Yoshinori Matsuo.
\newblock {Quantum focusing conjecture and the Page curve}.
\newblock {\em JHEP}, 12:050, 2023.

\bibitem{Anand:2023ozw}
Ankit Anand.
\newblock {Island in Warped AdS Black Holes}.
\newblock arXiv:2308.05432.

\bibitem{Li:2023zgy}
Pan Li and Yi~Ling.
\newblock {Refined symmetry-resolved Page curve and charged black holes*}.
\newblock {\em Chin. Phys. C}, 48(5):053109, 2024.

\bibitem{Xu:2023fad}
Yuanceng Xu, Dong Wang, and Qiyuan Pan.
\newblock {Page Curves in Holographic Superconductors}.
\newblock arXiv:2311.13145.

\bibitem{Ageev:2023hxe}
Dmitry~S. Ageev, Irina~Ya. Aref'eva, and Timofei~A. Rusalev.
\newblock {Black Holes, Cavities and Blinking Islands}.
\newblock arXiv:2311.16244.

\bibitem{Bousso:2023kdj}
Raphael Bousso and Geoff Penington.
\newblock {Islands Far Outside the Horizon}.
\newblock arXiv:2312.03078.

\bibitem{Jeong:2023lkc}
Hyun-Sik Jeong, Keun-Young Kim, and Ya-Wen Sun.
\newblock {Entanglement entropy analysis of dyonic black holes using doubly holographic theory}.
\newblock {\em Phys. Rev. D}, 108(12):126016, 2023.

\bibitem{Yadav:2022fmo}
Gopal Yadav.
\newblock {Page curves of Reissner\textendash{}Nordstr\"om black hole in HD gravity}.
\newblock {\em Eur. Phys. J. C}, 82:904, 2022.

\bibitem{Yadav:2022jib}
Gopal Yadav and Nitin Joshi.
\newblock {Cosmological and black hole islands in multi-event horizon spacetimes}.
\newblock {\em Phys. Rev. D}, 107(2):026009, 2023.

\bibitem{Yadav:2023sdg}
Gopal Yadav and Hemant Rathi.
\newblock {Yang-Baxter deformed wedge holography}.
\newblock {\em Phys. Lett. B}, 852:138592, 2024.

\bibitem{Yadav:2022mnv}
Gopal Yadav and Aalok Misra.
\newblock {Entanglement entropy and Page curve from the M-theory dual of thermal QCD above Tc at intermediate coupling}.
\newblock {\em Phys. Rev. D}, 107(10):106015, 2023.

\bibitem{RoyChowdhury:2022awr}
Anirban Roy~Chowdhury, Ashis Saha, and Sunandan Gangopadhyay.
\newblock {Role of mutual information in the Page curve}.
\newblock {\em Phys. Rev. D}, 106(8):086019, 2022.

\bibitem{Bhattacharya:2021dnd}
Aranya Bhattacharya, Arpan Bhattacharyya, Pratik Nandy, and Ayan~K. Patra.
\newblock {Partial islands and subregion complexity in geometric secret-sharing model}.
\newblock {\em JHEP}, 12:091, 2021.

\bibitem{Bhattacharya:2021nqj}
Aranya Bhattacharya, Arpan Bhattacharyya, Pratik Nandy, and Ayan~K. Patra.
\newblock {Bath deformations, islands, and holographic complexity}.
\newblock {\em Phys. Rev. D}, 105(6):066019, 2022.

\bibitem{murata2006hawking}
Keiju Murata and Jiro Soda.
\newblock Hawking radiation from rotating black holes and gravitational anomalies.
\newblock {\em Physical Review D}, 74(4):044018, 2006.

\bibitem{1971JETPL..14..180Z}
Ya.~B. {Zel'Dovich}.
\newblock {Generation of Waves by a Rotating Body}.
\newblock {\em Soviet Journal of Experimental and Theoretical Physics Letters}, 14:180, August 1971.

\bibitem{1974ApJ...193..443T}
S.~A. {Teukolsky} and W.~H. {Press}.
\newblock {Perturbations of a rotating black hole. III. Interaction of the hole with gravitational and electromagnetic radiation.}
\newblock {\em \apj}, 193:443--461, October 1974.

\bibitem{Press:1972zz}
William~H. Press and Saul~A. Teukolsky.
\newblock {Floating Orbits, Superradiant Scattering and the Black-hole Bomb}.
\newblock {\em Nature}, 238:211--212, 1972.

\bibitem{Bardeen:1999px}
James~M. Bardeen and Gary~T. Horowitz.
\newblock {The Extreme Kerr throat geometry: A Vacuum analog of AdS(2) x S**2}.
\newblock {\em Phys. Rev. D}, 60:104030, 1999.

\bibitem{page1993average}
Don~N Page.
\newblock Average entropy of a subsystem.
\newblock {\em Physical review letters}, 71(9):1291, 1993.

\bibitem{Hayden:2007cs}
Patrick Hayden and John Preskill.
\newblock {Black holes as mirrors: Quantum information in random subsystems}.
\newblock {\em JHEP}, 09:120, 2007.

\bibitem{sekino2008fast}
Yasuhiro Sekino and Leonard Susskind.
\newblock Fast scramblers.
\newblock {\em Journal of High Energy Physics}, 2008(10):065, 2008.

\bibitem{Nian:2019buz}
Jun Nian.
\newblock {Kerr Black Hole Evaporation and Page Curve}.
\newblock arXiv:1912.13474.

\bibitem{Nian:2023xmr}
Jun Nian.
\newblock {Hawking Radiation, Entanglement Entropy, and Information Paradox of Kerr Black Holes}.
\newblock arXiv:2312.14287.

\end{thebibliography}

\end{document}